\documentclass[pre,superscriptaddress,showpacs,preprint]{revtex4}
\usepackage{amsmath,graphicx,amssymb,amsfonts}%,times
\begin{document}\title{Hydrodynamic Theory of Granular Solids:\\
Permanent, Transient  and Granular Elasticity}
\author{Yimin Jiang}
\affiliation{Theoretische Physik, Universit\"{a}t
T\"{u}bingen,72076 T\"{u}bingen, Germany} \affiliation{Central
South University, Changsha 410083, China}
\author{Mario Liu}
\affiliation{Theoretische Physik, Universit\"{a}t
T\"{u}bingen,72076 T\"{u}bingen, Germany}
\date{\today}
\begin{abstract}
Although fully elastic when static, granular media become
transiently elastic when being slowly sheared -- during
which both the elastic energy and stress relax. Starting
from this observation, we cogently derive the framework
for granular hydrodynamics, a set of differential
equations consistent with general principles of physics,
especially reversible and irreversible thermodynamics. In
addition, an expression for the granular elastic energy
is reviewed and further discussed.
\end{abstract}  \maketitle

\section{Introduction}

In granular media, although the grains roll and slide,
in addition to being compressed and sheared, only the
latter, the deformation of the grains, leads to
reversible energy storage that sustains a static,
elastic stress, while rolling and sliding heats up the
system. The granular strain field $\varepsilon_{ij}$,
therefore, has two contributions, an elastic one
$u_{ij}$ accounting for deformation of the grains, and
a plastic one $p_{ij}$ for the rest, where
$\varepsilon_{ij}=u_{ij}+p_{ij}$. The elastic energy
$w(u_{ij})$ is a function of $u_{ij}$, not of
$\varepsilon_{ij}$, and the elastic contribution to
the stress $\sigma_{ij}$ is given as
$\pi_{ij}(u_{ij})\equiv-\partial w/\partial u_{ij}$.
With $\sigma_{ij}=\pi_{ij}$ in statics, stress balance
$\nabla_j\sigma_{ij}=0$ may be closed with
$\pi_{ij}=\pi_{ij}(u_{ij})$ and uniquely determined
employing appropriate boundary
conditions~\cite{J-L,ge}. Because the plastic part of
the strain needed for arriving at a given stress state
is quite irrelevant for its determination, one may
with certain justification consider static granular
media, say a sand pile at rest, as fully elastic.

If this sand pile is perturbed by periodic tapping,
circumstances change qualitatively. Its conic form will
then degrade until the surface becomes flat. This is
because part of the grains in the pile loose contact with
each other temporarily, during which their deformation
decreases. This implies a relaxing elastic strain
$u_{ij}$, and correspondingly, smaller elastic energy
$w(u_{ij})$ and static stress $\pi_{ij}(u_{ij})$. Since
the sand pile is no longer able to sustain static
stresses, it is now a transiently elastic system, same as
polymers -- though the respective microscopic mechanisms
are of course very different: temporary unjamming and
rearrangement of the grains versus slow disentanglement
of polymer strands. Note that flattening a sand pile
implies sizable granular rearrangement, requiring a
considerable portion of plastic strain $p_{ij}$.

Quantifying the random motion of the grains as granular
temperature $T_g$, we may take the relaxation time $\tau$
of the elastic strain $u_{ij}$ as a function of $T_g$,
with $\tau(T_g)\to\infty$ for $T_g\to0$. For vanishing
$T_g$, there is no strain relaxation, the deformation of
the grains are maintained, the sand pile keeps its conic
shape, and the system is elastic. For finite $T_g$, the
elasticity turns transient, with $u_{ij}$,
$\pi_{ij}(u_{ij})$ and $w(u_{ij})$ relaxing.

When granular media are being slowly sheared,
circumstances are similar. In addition to moving with the
large scale velocity $v_i$, the grains also move and slip
in deviation of it. This allows temporary, partial
unjamming, and implies a finite $T_g$, both again lead to
transient elasticity. Since $T_g$ is not always an
externally imposed parameter, as with tapping, but
frequently internally produced, especially by shear
flows, it is an independent variable of the granular
hydrodynamic theory, to be accounted for by its own
equation of motion. More specifically, the production of
$T_g$ by shear flows should have great similarities to
viscous heat production in normal fluids.

Granular media has different phases that, in dependence
of the grain's ratio of deformation to kinetic energy,
may loosely be referred to as gaseous, liquid and solid.
Moving fast and being free most of the time, the grains
in the gaseous phase have much kinetic, but next to none
elastic, energy~\cite{Haff}. In the denser liquid phase,
say in chute flows, there is less kinetic energy, more
deformation, and a rich rheology that has been
scrutinized recently~\cite{chute}. In granular statics,
with the grains deformed but stationary, the energy is
all elastic. This state is legitimately referred to as
solid because static shear stresses are sustained. If
granular solid is slowly sheared, the predominant part of
the energy remains elastic. As discussed, the system is
transiently elastic, or quasi-solid. In this paper, we
focus on the last two cases, and for simplicity refer to
both as the solid granular phase.

The transition between permanent and transient elasticity
is a crucial key to understanding granular solids. And
remarkably, it is as input quite sufficient for a formal
and cogent derivation of the framework for granular solid
hydrodynamics -- if one takes careful notice of all
general principles of physics, especially symmetry and
thermodynamic considerations. This is the first part of
the present paper. The second part deals with an concrete
expression for the granular elastic energy, how this
expression is supported by extensive experimental data
from granular statics. This is important because general
principles only confines the {\em structure} of the
hydrodynamic theory -- they yield a framework into which
many different theories fit. The three sets of
differential equations given below need the input of
specific expressions for the thermodynamic energy and the
transport coefficients. Only when their functional
dependence on the thermodynamic variables is given, do
the theories attain predictive power.

In the following, we first recall the hydrodynamic
theory of permanent and transient elasticity, in
\S~\ref{pe-GSH} and \S~\ref{te-GSH}; then merge both
to form granular hydrodynamics, in \S~\ref{ge-GSH}.
All equations in these three subsections are valid
irrespective what form the energy $w$ has. A specific
energy density suitable for granular media is then
reviewed and further discussed in \S~\ref{energy-GSH}.

In an accompanying paper~\cite{JL3}, we compare
hypoplasticity~\cite{Kolym}, a state of the art
engineering model on the behavior of granular solids,
with granular solid hydrodynamics as derived here, and
specified using the elastic energy of
\S~\ref{energy-GSH}.

\section{Elasticity Theory}
\subsection{Permanent Elasticity\label{pe-GSH}}
The conserved, thermodynamic energy density $w$ of
solids is a function of the symmetric strain field
$u_{ij}=u_{ji}$, and of the densities of entropy $s$,
mass $\rho$, momentum $\boldsymbol g$. So we write
(neglecting gravity)
\begin{equation}
\label{1-GSH}
 {\rm d}w = T{\rm d} s + \mu {\rm d}\rho + v_{i} {\rm d}
 g_{i} - \pi_{ij} {\rm d} u_{ij},
\end{equation}
denoting $T(s,\rho,u_{ij})\equiv\partial w/\partial s$,
$\mu(s,\rho,g_i,u_{ij})\equiv\partial w/\partial\rho$,
$v_i\equiv\partial w/\partial g_i=g_i/\rho$,
$\pi_{ij}(s,\rho,u_{ij})\equiv-\partial w/\partial
u_{ij}$. The equations of motion for the energy and its
variables are
\begin{eqnarray}
\label{2-GSH} {\textstyle\frac\partial{\partial t}}
w+\nabla_iQ_i=0, \qquad
{\textstyle\frac\partial{\partial t}}
s+\nabla_if_i=R/T,
\\
\label{3-GSH} {\textstyle\frac\partial{\partial t}}
\rho+\nabla_ij_i=0, \qquad
{\textstyle\frac\partial{\partial t}}
g_i+\nabla_j\sigma_{ij}=0,
\\
\label{4-GSH} {\textstyle\frac{\rm d}{{\rm d}
t}}u_{ij}-v_{ij}+[{\textstyle\frac12}\nabla_iy_j
+u_{ik}\nabla_{j}v_k+i\leftrightarrow j]=0,
\end{eqnarray}
where ${\textstyle\frac{\rm d}{{\rm d}
t}}={\textstyle\frac\partial{\partial t}}+v_k\nabla_k$
and $v_{ij}\equiv\frac12(\nabla_iv_j+\nabla_jv_i)$.
Expressing conservation laws and entropy production, the
first four equations are quite general and shared by all
hydrodynamic theories. Alone, they describe normal fluids
and represent the simplest hydrodynamic theory. The fifth
equation is characteristic of elastic systems, especially
ones that break the translational symmetry spontaneously.
(More on why Eq~(\ref{4-GSH}) must have the above form is
given in~\cite{temmen}.) Inserting
Eqs~(\ref{2-GSH}-\ref{4-GSH}) into the temporal
derivative of Eq~(\ref{1-GSH}),
\begin{equation}\label{5-GSH}
\textstyle\frac\partial{\partial t}w =
T\frac\partial{\partial t} s + \mu \frac\partial{\partial
t}\rho + v_{i}\frac\partial{\partial t} g_{i} - \pi_{ij}
\frac\partial{\partial t} u_{ij},
\end{equation}
and introducing the notations: $f^D_i$, $\sigma^D_{ij}$,
taking them to be given as
\begin{eqnarray}\label{6-GSH}
f_i&\equiv& sv_i-f^D_i,\\ \sigma_{ij}&\equiv&
\pi_{ij}-\pi_{ik}u_{jk}-\pi_{jk}u_{ik}\nonumber\\&&\quad
+(Ts+v_ig_i+\mu\rho+g_iv_j-w)-\sigma^D_{ij},
\label{7-GSH}
\end{eqnarray}
we obtain
\begin{eqnarray}\label{8-GSH}
\nabla_iQ_i=\nabla_i(Tf_i+\mu j_i
+v_j\sigma_{ij}-y_j\pi_{ij})\\\nonumber
+f_i^D\nabla_iT+\sigma_{ij}^Dv_{ij}
+y_i\nabla_j\pi_{ij}-R.
\end{eqnarray}
Clearly, one can write the left hand side of
Eq~(\ref{5-GSH}) as the divergence of something, plus
something else that vanishes in equilibrium (because the
so-called thermodynamic forces, $\nabla_iT, v_{ij}$ and
$\nabla_j\pi_{ij}$ do). Therefore, an inviting
possibility is to identify the first with the energy flux
$Q_i$, and the second with the entropy production $R$, a
quantity that also vanishes in equilibrium,
\begin{eqnarray}\label{9-GSH}
Q_i=Tf_i+\mu j_i
+v_j\sigma_{ij}-y_j\pi_{ij},\\\label{10-GSH}
R=f^D_i\nabla_iT+\sigma^D_{ij}v_{ij}+y_i\nabla_j\pi_{ij}.
\end{eqnarray}
This identification is in fact unique. It is easy to
verify that, as long as the energy $w$ remains general,
unspecified, there is no other way to write the left hand
side of Eq~(\ref{5-GSH}) as the sum of a divergence and
an expression that vanishes in equilibrium.

Taking in Eq~(\ref{10-GSH}) $(\nabla_iT, v_{ij},
\nabla_j\pi_{ij})$ as the thermodynamic forces, $(f^D_i,
\sigma^D_{ij}, y_i)$ as the fluxes, and forming each into
a 12-component vector, $\vec Y$ and $\vec Z$, the Onsager
force-flux relation gives their linear connection as,
\begin{equation}\label{10a-GSH}
\vec Z=\hat c\cdot\vec Y,
\end{equation}
where $\hat c$ is the transport matrix, with diagonal
elements that are positive, and off-diagonal ones that
satisfy the Onsager reciprocity relation. The simplest
example for $\hat c$ has only diagonal elements, all
positive scalars,
\begin{eqnarray}
\label{11-GSH}f^D_i&=&\kappa\nabla_iT,
\\ \label{12-GSH}\sigma_{ij}^D&=&\zeta
v_{\ell\ell}\delta_{ij}+\eta v^0_{ij}, \\
 \label{13-GSH} y_i&=&\beta^{P}\nabla_j\pi_{ij}.
\end{eqnarray}
Accounting for heat conduction and viscous stress, the
first two equations are shared by all hydrodynamic
theory. (The superscript $^0$, here and below, denotes
the traceless part of a tensor, eg. $v^0_{ij}\equiv
v_{ij}-\frac13\delta_{ij}v_{\ell\ell}$.) The third
accounts for permeation and defect motion, and is
specific to elastic media~\cite{perm}, see
section~\ref{SolidCreep}.

All elements of the matrix $\hat c$, usually referred to
as transport coefficients, are functions of the
thermodynamic variables, $s, \rho, u_{ij}$, or
alternatively, of the conjugate variables, $T, \mu,
\pi_{ij}$. In the generally accepted and above employed
linear version of the Onsager relation, they do not
depend on thermodynamic forces, $\nabla_iT, v_{ij},
\nabla_j\pi_{ij}$. So we may take the coefficients
$\kappa, \eta,\zeta$ and $\beta^P$ to depend on the
temperature, the pressure, and scalar combinations of the
stress, such as $\pi_{\ell\ell}$ and
$\pi_s^2\equiv\pi_{ij}^0\pi_{ij}^0$.

\subsubsection{Solid Creep Motion\label{SolidCreep}}

Enforcing a steady velocity at the surface of granular
solid, the velocity field is observed to penetrate rather
deep into the bulk of the granular medium, with a
magnitude that decays exponentially with
depth~\cite{creep}. The usual collective modes of
velocities in hydrodynamic theories of elastic media are
of course such that they reduce to a constant velocity
when stationary (sound), or one that varies linearly in
space (shear diffusion). But there is also a less-known
one that decays exponentially, a consequence of
Eq~(\ref{13-GSH}) and the less studied permeation
coefficient $\beta^P$. We shall refer to this mode as
``solid creep motion."

Linearized with respect to velocity, Eq~(\ref{4-GSH})
reduces, for the stationary case $\partial
u_{ij}/\partial t=0$, to
\begin{equation}%\label{}
v_{ij}={\textstyle\frac12}
\beta^P\nabla_k(\nabla_i\pi_{jk}+\nabla_j\pi_{ik}),
\end{equation}
implying that mass and shear flows are possible without
any changes in the elastic strain field, or equivalently,
in the elastic stress and elastic energy. Similarly,
momentum conservation, or Eq~(\ref{3-GSH}), linearized
and in steady flow, $\partial(\rho v_i)/\partial t=0$,
reduces to
\begin{equation}%\label{}
\nabla_j(D\delta_{ij}+\pi_{ij}-\eta v^0_{ij})=0
\end{equation}
(where $D\delta_{ij}$ stands for the diagonal terms that
do not concern us here). Now, consider a half space $y>0$
filled with solid, which has its surface at $y=0$, moving
with a given velocity along $x$. Permitting only a
$y$-dependence in this one-dimensional geometry, we have
\begin{equation}%\label{}
v_{xy}={\textstyle\frac12}\beta^P\nabla^2_y\pi_{xy},\quad
\nabla_y(\pi_{xy}-\eta v^0_{xy})=0.
\end{equation}
These two equations clearly imply exponentially
decaying velocity $v_x$ and change of the elastic
stress $\delta\pi_{xy}$,
\begin{equation}\label{alpha}
v_x,
\delta\pi_{xy}\sim\exp\frac{-y}{\sqrt{{\textstyle\frac12}\eta\,\beta^P}}.
\end{equation}

In granular medium, this behavior will be modified,
because the elasticity there is transient rather than
permanent. But should solid creep motion retains its
qualitative behavior under certain circumstances,
Eq~(\ref{alpha}) would constitute a natural explanation
of granular creep flow.

\subsection{Transient Elasticity\label{te-GSH}}

Although the equations of the last section are fairly
general and account for all kinds of elasticity, linear
as well as nonlinear, they do exclude transient
elasticity, such as realized in polymers. In these,+
elasticity arises from entanglement of polymer strands,
which are stretched and sheared, if not given enough time
to disentangle. But if given enough time, the
deformation, with it also the associated energy and
stress, relax. So the system is to be accounted for by a
set of equations which reduce to those of the last
section for small time spans, but allow the deformation
$u_{ij}$ to relax for longer time spans.

The independent variables remain the same, so do the
conservation laws. So
Eqs~(\ref{1-GSH},\ref{2-GSH},\ref{3-GSH}) are unchanged,
but Eq~(\ref{4-GSH}) is modified to allow for a
relaxation term $X_{ij}$
\begin{equation}\label{15-GSH}
{\textstyle\frac{\rm d}{{\rm d} t}}
u_{ij}-v_{ij}+[{\textstyle\frac12}\nabla_iy_j
+u_{ik}\nabla_{j}v_k+i\leftrightarrow j]=X_{ij}.
\end{equation}
The same calculation of Eq~(\ref{5-GSH}), with the same
notation of Eq~(\ref{6-GSH},\ref{7-GSH}), then leads to
the same energy flux $Q_i$, but a modified entropy
production,
\begin{equation}\label{16-GSH}
R=f^D_i\nabla_iT+\sigma^D_{ij}v_{ij}+y_i\nabla_j\pi_{ij}
+X_{ij}\pi_{ij}.
\end{equation}
This implies $\pi_{ij}$ is now not only a conjugate
variable, but also a thermodynamic force, increasing the
dimension of the 12-component vector $\vec Y$ in
Eq~(\ref{10a-GSH}) by another 6 components. Similarly,
the vector $\vec Z$ is also increased by the 6 components
of $X_{ij}$, and $\hat c$ is now a $18\times18$-matrix.
Other from that, Eq~(\ref{10a-GSH}) still holds. The
simplest, diagonal and scalar example is again given by
Eqs~(\ref{11-GSH}, \ref{12-GSH}, \ref{13-GSH}), in
addition to
\begin{equation}\label{17-GSH}
X_{ij}=\beta
\pi_{ij}^0+\beta_1\pi_{\ell\ell}\,\delta_{ij},
\end{equation}
a term that permits $u_{ij}$ to relax, as long as
$\pi_{ij}$ is nonzero.

As discussed in the last paragraph of \S~\ref{pe-GSH},
the transport coefficients $\beta, \beta_1$ are
functions of the thermodynamic variable $u_{ij}$, or
equivalently, of $\pi_{ij}=\pi_{ij}(u_{ij})$. This
remains true even though $\pi_{ij}$ is now also part
of $R$, Eq~(\ref{16-GSH}), and hence an additional
thermodynamic force.

A point worth clarifying concerns the  plastic strain
$p_{ij}$: The total strain $\varepsilon_{ij}=u_{ij}+
p_{ij}$, a purely kinematic quantity,  obeys the equation
\begin{equation}%\label{}
{\textstyle\frac{\rm d}{{\rm d} t}}\,\varepsilon_{ij}
+[\varepsilon_{ik}\nabla_{j}v_k+i\leftrightarrow
j]=v_{ij}.
\end{equation}
So, as a result of Eq~(\ref{15-GSH}), the plastic strain
is determined by
\begin{equation}%\label{}
{\textstyle\frac{\rm d}{{\rm d} t}}\,
p_{ij}+[-{\textstyle\frac12}\nabla_iy_j+p_{ik}\nabla_{j}v_k+i\leftrightarrow
j] =-X_{ij}.
\end{equation}
If a transiently elastic medium is quickly and uniformly
deformed, such that there is no time for relaxation,
$X_{ij}\approx0$, we have $\varepsilon_{ij}=u_{ij}$ after
the deformation. Holding it for a while, $v_{ij}=0$, the
elastic deformation $u_{ij}$ relaxes to zero, while the
plastic one $p_{ij}$ grows accordingly, until one
replaces the other completely, and we have
$p_{ij}=\varepsilon_{ij}$. The system now stays where it
is, and the initial displacement is referred to as
``plastic," rather than elastic, because it does not have
the tendency to return to the original position.

Essentially this set of equations, as specified in
this section, was recently shown well able of
accounting for the full range of polymers'
non-Newtonian behavior, including shear-thinning,
elongational strain-hardening, rod climbing (the
Weissenberg effect), and various empirical rules such
as Cox-Merz and First Gleissle Mirror
Rule~\cite{temmen,om}

\subsection{Granular Elasticity\label{ge-GSH}}

As discussed, sand and other granular media display both
elastic and transiently elastic behavior -- depending on
whether the granular temperature $T_g$ vanishes or not.
Including the density of granular entropy $s_g$ as an
additional, independent thermodynamic variables, the
Gibbs relation of Eq~(\ref{1-GSH}) now reads
\begin{equation}
\label{20-GSH}
 {\rm d}w = T{\rm d} s +T_g{\rm d}s_g
 + \mu {\rm d}\rho + v_{i} {\rm d}
 g_{i} - \pi_{ij} {\rm d} u_{ij}.
\end{equation}

Granular temperature is not a new concept. Haff, also
Jenkin and Savage~\cite{Haff}, were probably the first
to introduce it in the context of granular gas, using
it to denote the average kinetic energy of the grains.
Hence $T_g\sim\epsilon_k$, where $\epsilon_k$ is the
kinetic energy density. Nowadays, this $T_g$ is
routinely used in considering granular gas and
liquid~\cite{Lub}. Note that given this interpretation
of $T_g$, we have $T_g=\partial \epsilon_k/\partial
s_g\sim\partial T_g/\partial s_g$, and the granular
entropy is uniquely determined, $s_g\sim \ln T_g$.
More recently, there is much discussion of a
configurational entropy $S_c$ in the literature. The
original concept by Edwards was to approximate grains
as infinitely rigid and all configurations as having
identical energy~\cite{Edw}, so $S_c$ is a function
only of the system's volume. When relaxing the
rigidity approximation, and allowing the elastic
energy to vary, $S_c$ is again a function of energy
and volume $S_c=S_c(E,V)$, and a configurational
temperature is naturally given as $T_c^{-1}=\partial
S_c/\partial E$ (see~\cite{Nic} for a review).

In thermodynamics, the energy change ${\rm d}w$ from all
microscopic, implicit variables is subsumed as $T{\rm
d}s$, with $s$ the entropy and $T\equiv\partial
w/\partial s$ its conjugate variable. From this, we
divide out the kinetic energy of granular random motion,
executed by the grains in deviation from the ordered,
large-scale motion, denoting it as $T_g{\rm d}s_g$, and
calling $s_g$ and $T_g\equiv\partial w/\partial s_g$
granular entropy and temperature, respectively. In other
words, we consider two heat reservoirs, the first
containing the energy of granular random motion, the
second the rest of all microscopic degrees of freedom,
especially phonons. In equilibrium, $T_g=T$, and $s_g$ is
part of $s$. (In fact, we may simply forget $s_g$, since
it has far less degrees of freedom.) But when the
granular system is being tapped or sheared, and $T_g$ is
many orders of magnitude larger than $T$, then this
leaky, intermediary heat reservoir can no longer be
ignored. As $s_g$ then serves as a nonhydrodynamic,
macroscopically slow degree of freedom, with $T_g$ its
conjugate variable.

Taking $s_g$ as the part of the entropy accounting for
the granular kinetic energy, our definition is fairly
close to the entropy of granular gas discussed above,
as given by Haff, though its functional dependence
will probably be modified, because it must be
evaluated taking into consideration the effect of
excluded volumes -- an overwhelming one in the dense
solid phase [see Eq~(9) of the third of~\cite{Lub}].
The concept of configurational entropy, on the other
hand, is closer to our second heat reservoir, the true
entropy $s$, see section 6 of the first, and section
10 of the third, reference~\cite{ge}, for a discussion
of their relationship.

The functional dependence of $s_g(T_g)$, more precisely,
the equation of state $T_g=T_g(s,s_g,\rho,u_{ij})$, is
given once the energy $w$ is known. Although all
equations of this and the last two sections remain valid
irrespective of what special expression one chooses for
$w$, concrete predictions certainly depend on it. Since
it appears difficult, at least at present, to evaluate
$w$ microscopically, one may alternatively employ
experimental data in conjunction with general
considerations to narrow down its possibility. We shall
examine $w$'s dependence on $u_{ij}$ and $\rho$ in the
next section, but defer that on $s_g$ to a future
publication.

Taking the balance equation for $s_g$, in the uniform
case, as ${\frac\partial {\partial t}}s_g =R_g/T_g$, we
first of all need $R_g$ to contain the term $-\gamma
(T_g-T)^2$. This is because being a slow, nonhydrodynamic
variable, the equation of motion for $s_g$ should have
the usual relaxation form, ${\frac\partial {\partial
t}}s_g =-\gamma (T_g-T)$, pushing $T_g$ towards the
ambient temperature $T$. (Since any random motion of the
grains implies such improbably high $T_g$, neglecting $T$
in this expression is always an excellent approximation.
We shall therefore from here on always write
${\frac\partial {\partial t}}s_g =-\gamma T_g$.)

Second, with the heat bath divided into two parts,
viscous heat production should fill both baths
simultaneously. Therefore, we keep the term
$\sigma^D_{ij}v_{ij}$ in $R$, with $\sigma^D_{ij}=\eta
v_{ij}^0+\zeta v_{\ell\ell}$, and write the analogous
one, $\Sigma^D_{ij}v_{ij}$ into $R_g$, with
$\Sigma^D_{ij}=\eta_g v_{ij}^0+\zeta_g v_{\ell\ell}$
denoting the viscous stress contribution from exciting
granular random motion. The magnitude of the four
viscosities depend on microscopic details and cannot be
decided on general principles. For instance, while $\eta$
is probably a small quantity compared to $\eta_g$ in dry
sand, because macroscopic shear flows excite granular
random motion first, $\eta$ should be quite a bit larger
in wet sand: A macroscopic shear flow implies much
stronger microscopic shear flows in the fluid layers
between grains, and the energy dissipated in these layers
should predominantly go to $s$, rather than to $s_g$
first.

Third, granular entropy production $R_g$ should have the
term $\kappa_g\nabla_iT^2_g$, from an inhomogeneous
granular temperature, in exact analogy to the term
$\kappa\nabla_iT^2$ in $R$. So the final expression
should be $R_g=\Sigma^D_{ij}v_{ij}+ \kappa_g\nabla_iT^2_g
-\gamma T_g^2$. A direct and desirable consequence of
this expression is that for stationarity, ${\frac\partial
{\partial t}}s_g=R_g/T_g =0$, and a constant $T_g$, any
shear flows excite the granular temperature of $\gamma
T_g^2=\eta_g v_{ij}^0v_{ij}^0+\zeta_g v_{\ell\ell}^2$,
which is (as discussed) what renders granular elasticity
transient.

We do not have good reasons for ruling out a term in
$R_g$ analogous to $y_i\nabla_j\pi_{ij}$, or one
$\sim\nabla_iT^2_g$ in $R$. But neither is there any
experimental evidence demanding their existence. So
although both are allowed for the general case, they are
left out here for the simplicity of display. On the other
hand, a term in $R_g$ analogous to $X_{ij}\pi_{ij}$
cannot exist, because we would then have $\gamma
T_g^2=X_{ij}\pi_{ij}$ for granular statics, implying a
finite $T_g$ and decaying sand piles.

Given the above consideration specifying $R_g$, we may
embark on the derivation of the equations of motion for
granular elasticity, in the same way as above. We start
from the following equations,
\begin{eqnarray}
\label{2y-GSH} {\textstyle\frac\partial{\partial t}}
w+\nabla_iQ_i=0, \quad {\textstyle\frac\partial{\partial
t}} \rho+\nabla_ij_i=0,\qquad
\\ \label{3y-GSH}
{\textstyle\frac\partial{\partial t}}
s+\nabla_if_i=R/T,\quad {\textstyle\frac\partial{\partial
t}} s_g+\nabla_iF_i=R_g/T_g,
\\
{\textstyle\frac\partial{\partial t}} g_i+\nabla_j
\sigma_{ij}=0,\qquad\qquad\qquad\label{3yA-GSH}
\\
\label{4y-GSH} {\textstyle\frac{\rm d}{{\rm d} t}}
u_{ij}-v_{ij}+[{\textstyle\frac12}\nabla_iy_j
+u_{ik}\nabla_{j}v_k+i\leftrightarrow j]=X_{ij}.
\end{eqnarray}
Inserting these into Eq~(\ref{20-GSH}),
\begin{equation}\label{5y-GSH}
\textstyle\frac\partial{\partial t}w =
T\frac\partial{\partial t} s +T_g\frac\partial{\partial
t} s_g + \mu \frac\partial{\partial t}\rho +
v_{i}\frac\partial{\partial t} g_{i} - \pi_{ij}
\frac\partial{\partial t} u_{ij},
\end{equation}
using the notations
\begin{eqnarray}\label{6y-GSH}
f_i\equiv sv_i-f^D_i,\quad F_i\equiv
s_gv_i-F^D_i,\qquad\qquad\quad
\\\nonumber
\sigma_{ij}\equiv(-w+Ts+v_ig_i+\mu\rho+T_gs_g)
\delta_{ij}\qquad\qquad\\
+\pi_{ij}-\pi_{ik}u_{jk}-\pi_{jk}u_{ik}+g_iv_j-\sigma^D_{ij}-\Sigma^D_{ij},
\label{7y-GSH}
\end{eqnarray}
we obtain
\begin{eqnarray}\label{8y-GSH}
\nabla_iQ_i=\nabla_i(Tf_i+T_gF_i+\mu j_i
+v_j\sigma_{ij}-y_j\pi_{ij})\qquad\quad \\\nonumber
-R+f_i^D\nabla_iT +y_i\nabla_j\pi_{ij}%\\\nonumber
+\sigma_{ij}^Dv_{ij}+X_{ij}\pi_{ij}+\gamma T_g^2
\\\nonumber
-R_g+\Sigma_{ij}^Dv_{ij}+F_i^D\nabla_iT_g -\gamma T_g^2
\end{eqnarray}
and deduce
\begin{eqnarray}\label{9y-GSH}
Q_i&=&Tf_i+T_gF_i+\mu j_i +v_j\sigma_{ij}-y_j\pi_{ij},
\\\nonumber
R&=&f_i^D\nabla_iT +y_i\nabla_j\pi_{ij}\\
&&\qquad\quad+\sigma_{ij}^Dv_{ij}+X_{ij}\pi_{ij}+\gamma
T_g^2,\label{10y-GSH}\\\label{11y-GSH}
R_g&=&\Sigma_{ij}^Dv_{ij}+F_i^D\nabla_iT_g -\gamma T_g^2.
\end{eqnarray}
Given the expressions for $R$, we may take flux vector as
$\vec Z=(f^D_i, y_i, \sigma^D_{ij}, X_{ij})$, the force
vectors as $\vec Y=(\nabla_iT, \nabla_j\pi_{ij}, v_{ij},
\pi_{ij})$, and again formulate the Onsager force-flux
relation as $\vec Z=\hat c\cdot\vec Y$. Analogously,
given $R_g$, we have $\vec Z_g=\hat c_g\cdot\vec Y_g$,
where $\vec Z_g=(F^D_i, \Sigma^D_{ij})$ and $\vec
Y_g=(\nabla_iT_g, v_{ij})$. In  addition, we require
\begin{equation}%\label{}
X_{ij}\to0\quad \text{for}\quad T_g\to0,
\end{equation}
to ensure permanent elasticity in granular statics.

This completes the derivation and presentation of the
structure of a hydrodynamics of permanent elasticity at
$T_g=0$, and transient elasticity at finite $T_g$. To
find Granular solid hydrodynamics, we still need to
specify the energy $w$, and the functional dependence of
the transport matrices, $\hat c, \hat c_g$. Instead of a
microscopic derivation of these quantities starting from
some specific interaction, we employ general
considerations (such as requiring $w$ to have a positive
curvature where the system is stable, see \S~\ref{yield})
and experimental data to narrow down the possibilities.
Hereby, $w$ may be determined by static data alone, but
$\hat c, \hat c_g$ must be considered using data from
granular dynamics. The simplest example is again given by
$\hat c, \hat c_g$ being both diagonal,
\begin{eqnarray}
\label{11y-GSHy}f^D_i=\kappa\nabla_iT,\qquad%\qquad
F^D_i=\kappa_g\nabla_iT_g,\quad
y_i=\beta^{P}\nabla_j\pi_{ij},
\\
\label{12y-GSHy} \Sigma_{ij}^D=\zeta_g
v_{\ell\ell}\delta_{ij}+\eta_g v^0_{ij},\quad
\sigma_{ij}^D=\zeta v_{\ell\ell}\delta_{ij}+\eta
v^0_{ij},
\\X_{ij}=\beta\pi_{ij}^0
+\beta_1\delta_{ij}\pi_{\ell\ell}.\qquad\qquad
\end{eqnarray}

In the next section, \S~\ref{energy-GSH}, an energy
expression appropriate for granular media is presented,
and shown to account for important features of granular
statics. For the homogeneous case, with $\nabla_iT,
\nabla_iT_g, \nabla_j\pi_{ij}=0$, we propose to combine
this $w$ with the following transport structure, diagonal
except for the two terms preceded by $\alpha$,
\begin{eqnarray}\sigma_{ij}^D+
\Sigma_{ij}^D&=&(\zeta+\zeta_g)
v_{\ell\ell}\delta_{ij}+(\eta+\eta_g)
v^0_{ij}+\alpha\pi_{ij},\quad \label{13y-GSH}
\\\label{14y-GSH}
X_{ij}&=&-\alpha v_{ij}-\frac{u _{ij}^0}\tau
-\frac{u_{\ell\ell}\,\delta_{ij}}{\tau_1}.
\end{eqnarray}
The first equation is simply a sum of the two
dissipative stress contributions. The second equation
uses the specific form of $w$, a result of which is
\begin{equation}%\label{}
\pi_{ij}\equiv-\frac{\partial w}{\partial u_{ij}}
=\sqrt\Delta({\cal B}\Delta\,\delta _{ij}-2{\cal A}\,
u_{ij}^0) +{\cal A} \frac{u_s^2}{2\sqrt\Delta}\delta
_{ij},
\end{equation}
see Eq~(\ref{8}) below. So the relaxation times are
given as
\begin{equation}\label{GSH1}
\frac1\tau\equiv2\beta{\cal A}\sqrt\Delta, \quad
\frac1{\tau_1}\equiv3\beta_1\sqrt\Delta\left({\cal
B}+\frac{{\cal A}u_s^2}{2\Delta^2}\right).
\end{equation}
Obviously, a simplification is given by taking either
$\beta$ and $\beta_1$, or $\tau$ and $\tau_1$, as
independent from $u_{ij}$. Choosing the second
possibility, and taking $\tau,\tau_1$ as proportional to
$T_g$, all other coefficients (ie. $\zeta,\zeta_g,
\eta,\eta_g,\alpha$) as constant gives us a complete and
well specified theory. As will be shown in an
accompanying paper~\cite{JL3}, this choice leads to a
surprisingly good agreement with
hypoplasticity~\cite{Kolym}, a modern engineering theory
widely employed to model solid granular behavior,
especially triaxial experiments.

\subsubsection{Granular Gas}
Since we are considering a hydrodynamic theory, we
should expect the equations as given above to easily
connect to that of granular gas, such as given
in~\cite{Haff} by Haff. Taking the elastic strain to
relax infinitely fast, $\tau,\tau_1\to0$, essentially
eliminates $u_{ij}$ as an independent variable. As a
result, we have $w=w(T,T_g,\rho)$ in the rest frame,
and only Eqs~(\ref{2y-GSH},\ref{3y-GSH},\ref{3yA-GSH})
remain as equations of motion, with the dissipative
currents given by the second of Eqs~(\ref{11y-GSHy}),
and the first of Eqs~(\ref{12y-GSHy}). Following Haff,
we may take $w\sim T_g$, $s_g\sim\ln T_g$, and the
term $(T_gs_g+\mu\rho-w)\,\delta_{ij}$ as the main
contribution to the pressure [see Eq~(\ref{7y-GSH})];
also
\begin{equation}%\label{}
\zeta_g,\,\eta_g,\, \kappa_gT_g,\, \gamma T_g
\sim\,\rho \sqrt{T_g}.
\end{equation}
%$$.
[Because $s$ is not included as an
independent variable, the first of Eqs~(\ref{3y-GSH}),
is ignored in~\cite{Haff}, as are $\kappa,
\eta,\zeta$. Moreover, $\Sigma_{ij}^D$ are included
only in $R_g$, not in the stress flux $\sigma_{ij}$,
which is perhaps not quite consistent. The general
gist, however, is certainly the same.]

\section{A Granular Energy Expression\label{energy-GSH}}
Linear elasticity is a simple, consistent and complete
theory. It starts with an energy $w$ that depends on the
strain, $u_{ij}=\frac12(\nabla _iU_j+\nabla _jU_i)$, with
$U_i$ the displacement vector,
\begin{equation}\label{1}
w=\textstyle\frac12K\Delta^2+\mu u_s^2\quad (\Delta\equiv
-u_{\ell\ell},\, u_s\equiv \sqrt{u^0_{ij}u^0_{ij}}),
\end{equation}
see~\cite{LL7}. $K,\mu>0$ are two material-dependent
constants, referred to as the bulk and shear modulus.
($u_{\ell\ell}$ is the trace of $u_{ij}$, and
$u^0_{ij}\equiv u_{ij}-\frac13u_{\ell\ell}\,\delta_{ij}$
its traceless part.) The stress-strain relation is
obtained as a derivative,
\begin{equation}\label{2}
\sigma_{ij}=\pi_{ij}\equiv-\frac{\partial w}{\partial
u_{ij}}=K\Delta\,\delta_{ij}-2\mu\, u^0_{ij},
\end{equation}
which contains the pressure $P$ and the scalar shear
stress $\sigma_s$,
\begin{equation}\label{3}
P\equiv\textstyle\frac13\sigma_{\ell\ell}=K\Delta,\quad
\sigma_s\equiv\sqrt{\sigma^0_{ij}\sigma^0_{ij}}=2\mu u_s,
\end{equation}
both employed frequently below. Note that as there is
no difference between $\sigma_{ij}$ and $\pi_{ij}$ in
statics, we shall use them interchangeably here, in
\S~\ref{energy-GSH}.

Some ramifications of linear elasticity are: (1)~Since
the stress $\sigma_{ij}$ is given as a function of
three variables, $U_i$, the three components of the
force balance $\nabla_j\sigma_{ij}=\rho G_i$ (with
$\rho$ the density and $G_i$ the gravitational
constant) suffice to uniquely determine $U_i$, from
which the stress $\sigma_{ij}$ may be calculated for
arbitrary geometry. (2)~The inverse compliance tensor,
$M_{ijk\ell}$, linking the increments of stress and
strain, $\text{d}\sigma_{ij}$ and $\text{d}u_{k\ell}$,
is both isotropic and constant,
\begin{eqnarray}
  \label{4}
\text{d}\sigma_{ij}=\frac{\partial\sigma_{ij}}{\partial
u_{k\ell}}\,\text{d} u_{k\ell}\equiv
M_{ijk\ell}\,\text{d} u_{k\ell},\\
M_{ijk\ell}=K\delta_{ij}\delta_{k\ell}-
\mu(\delta_{ik}\delta_{j\ell}+\delta_{jk}\delta_{i\ell}).
\label{5}
\end{eqnarray}
(3)~As the pressure $P=K\Delta$ does not depend on the
shear $u_s$, there is no volume dilatancy, $(\partial
P/\partial u_s)|_\Delta=0$. (4)~Yield is not predicted.
[Note that while the points (2), (3), (4) depend on the
form of the energy $w$, the statement under (1) is quite
general.]

These equations account well for ordinary solids, but not
for granular systems. Sand displays volume dilatancy,
possesses a compliance tensor with significant
stress-induced anisotropy, and most importantly, never
strays far from yield, displaying significant
irreversible, fluid-like, plastic movements in its
vicinity.

The first attempt to modify linear elasticity, so as
to better account for granular behavior, was due to
Boussinesq~\cite{Gudehus}. He assumed, around 1874,
stress-dependent elastic moduli, $K,\mu\sim \Delta
^{1/2}\sim P^{1/3}$, in Eq~(\ref{2}),
\begin{equation}\label{6}
\sigma _{ij}\sim\sqrt\Delta\left( \Delta\, \delta
_{ij}-\frac{3-6\nu }{1+\nu }\,u_{ij}^0\right),\quad
\frac{3-6\nu }{1+\nu }=\frac{2\mu}K,
\end{equation}
with $\nu$ the constant Poisson ratio. This nonlinear
stress-strain relation, sometimes referred to as the
``quasi-elastic model," is employed to understand
granular compression~\cite{Evesque-de-Gennes} and
sound velocity~\cite{Goddard}. Unfortunately, the
above failure list of linear elasticity remains partly
intact: \textbullet~As $P$ remains a function of
$\Delta$ alone, dilatancy vanishes, $\partial
P/\partial u_s|_\Delta=0$. \textbullet~Yield must
still be postulated. In addition, Eq~(\ref{6})
contains a basic deficiency: No energy $w$ exists such
that $\sigma_{ij}=-\partial w/\partial u_{ij}$ holds,
because the associated Maxwell relation is violated,
$\partial \sigma _{ij}/\partial u_{\ell k}\not
=\partial \sigma _{\ell k}/\partial u_{ij}$.

We choose the granular elastic energy to be~\cite{J-L}
\begin{equation}
w=\sqrt\Delta\left( \textstyle\frac 25{\cal B}\Delta
^2+{\cal A}u_s^2\right), \label{7}
\end{equation}
with ${\cal A,B}>0$ denoting two material constants. The
associated stress is
\begin{equation}
\sigma _{ij}=\sqrt\Delta({\cal B}\Delta\,\delta
_{ij}-2{\cal A}\, u_{ij}^0) +{\cal A} \frac
{u_s^2}{2\sqrt\Delta}\delta _{ij}. \label{8}
\end{equation}
As compared to Eq~(\ref{6}), the only difference is the
last term $\sim u_s^2/\sqrt\Delta$. This is, however,
amazingly useful in accounting for granular behavior. As
we shall see, it yields volume dilatancy, shear-induced
anisotropy, and above all, predicts yield at the Coulomb
condition,
\begin{equation}
\sigma _s/P=\sqrt{2{\cal A/B}}.  \label{9}
\end{equation}

In granular materials, there is a regime in which
dissipation is insignificant and elastic responses
dominant: small-amplitude perturbations from given points
in the stress space. This is shown by Kuwano and
Jardine~\cite{Kuwano-Jardine} experimentally, who
observed that stress increments become reversible if the
strain fluctuations are around $10^{-4}$. It is also
corroborated by Alonso-Marroquin and Herrmann~\cite{AH}
in molecular-dynamic simulations: Reducing elastic
strains to $10^{-6}$, the irreversible plastic
contributions are found around $10^{-14}$, implying a
line as the stress-strain response, rather than the usual
ellipse at higher amplitudes.

This fact is important because it makes a direct
verification of Eq~(\ref{8}) possible: Measure
$\text{d}\sigma_{ij}= ({\partial\sigma_{ij}}/{\partial
u_{k\ell}})\,\text{d} u_{k\ell}$ and $\text{d} u_{k\ell}$
independently, and compare the result to
$M_{ijk\ell}\equiv \,{\partial\sigma_{ij}}/{\partial
u_{k\ell}}$ as calculated from Eq~(\ref{8}). The data
in~\cite{Kuwano-Jardine} are extensive, comprising of 36
independent components of $M_{ijk\ell}$, all as functions
of pressure, shear and the void ratio $e$. Comparing
these data to the calculate $M_{ijk\ell}$ is the main
result of this section, and represents an ambitious test
of the energy $w$, Eq~(\ref{7}): Energy and stress of
Eqs~(\ref{7},\ref{8}) depend only on two material
parameters, $\cal A$ and $\cal B$, with their ratio fixed
by the yield condition, Eq~(\ref{9}). Since the Ham river
sand used in the experiment has a Coulomb yield angle of
around $28^\circ$, implying $\xi\equiv{\cal B/A}=5/3$,
only $\cal A$, a scale factor and a measure of the total
hardness, is left as an adjustable parameter. Taking
${\cal A}=5100$~Mpa, we find satisfactory agreement with
their data at all values of pressure and shear, for the
void ratio $e=0.66$ --- except close to yield which, due
to increased plastic contributions, represents an
especially difficult experimental regime. Because Kuwano
and Jardine noticed that $e$ only alters the total
hardness, by the factor $f\equiv(2.17-e)^2/(1+e)$, taking
${\cal A,B}\sim f$ achieves agreement with respect to any
other values of $e$ as well. Similar agreement to their
data on ballotini (glass beads) was achieved by taking
${\cal A}=4200$~Mpa. Therefore, we take
\begin{equation}
\mathcal{A}= \mathcal{A}_0\times\frac{(2.17-e)
^2}{1.3736(
1+e)},\quad \xi\equiv{\cal\frac BA}=\frac53 %\label{Hardin-Richart}
\end{equation}
with $\mathcal{A}_0=5100$ and 4200~Mpa being the value of
$\cal A$ for $e=0.66$, for Ham river sand and ballotini,
respectively.

Given this experimental support on the functional
dependence of $\sigma_{ij}$ on $U_k$, we have employed
Eq~(\ref{8}) to evaluate static stress distributions in
silos, sand piles and under point loads, not surprisingly
with rather satisfactory results, see~\cite{ge}. Note
that Eq~(\ref{8}) does not contain any fit parameters:
$\xi=5/3$ is fixed by the yield angle, while ${\cal
A}_0$, as a scale factor, does not enter the stress
distribution at all. (Given a solution, one may change
the strain by the factor $\alpha$,  and ${\cal A}_0$ by
$\alpha^{-1.5}$, with the stress unchanged and still a
solution, provided the boundary conditions are the usual
ones, either given in terms of stresses or require that
the displacement vanishes.)

%%%%%%%%%%%%%%%%%%%%%%%%%%%%%%%%%%%%%%%%%%%%%%%%%%%%%

\subsection{Yield and Energetic
Instability}\label{yield}

A thermodynamic energy must be a convex function of state
variables to ensure stability -- this is why
compressibility and specific heat are always positive,
cf.~\cite{Callen}. Being a quadratic function of $\Delta$
and $u_s$, the energy of linear elasticity, Eq~(\ref{1}),
is always convex. Conversely, the granular energy,
Eq~(\ref{7}), is convex if and only if
\begin{eqnarray}
\left( \partial ^2w/\partial \Delta ^2\right) _{u_s}
&\geq &0,\ \ \left(
\partial ^2w/\partial u_s^2\right) _\Delta \geq 0,
\label{10} \\ \left( \partial ^2w/\partial \Delta
\partial u_s\right) ^2 &\leq &\left(
\partial ^2w/\partial \Delta ^2\right) _{u_s}
\left( \partial ^2w/\partial u_s^2\right) _\Delta
\label{11}
\end{eqnarray}
hold. (See appendix on some subtleties in this context.)
More explicitly, this implies
\begin{equation}
u_s^2/\Delta ^2\leq 2{\cal B/A},  \label{12}
\end{equation}
drawing the boundary for the region of stable strains.
Deriving $4P/\sigma _s=(\Delta /u_s)\times \left( 2{\cal
B}/{\cal A}+u_s^2/\Delta ^2\right)$ from Eq~(\ref {8}),
and inserting $u_s^2/\Delta ^2= 2{\cal B/A}$ into it,
Eq~(\ref{9}), the Drucker-Prager version of the Coulomb
yield condition (cf. Schofield \& Wroth, 1968; Huang,
1983)  is obtained. The actual Coulomb yield condition,
$\sigma _s/P=(\sqrt{18+6L^2}\sin \varphi _c)/({L\sin
\varphi _c+3})$, where $L\equiv\sqrt{3}\tan \left[ \frac
13\arcsin \left( \sqrt{6}\,\sigma _{ij}^0\sigma
_{jk}^0\sigma _{ki}^0/\sigma _s^3\right) \right] $
denotes the Lode parameter, would only result if terms
$\sim u_{ij}^0u_{jk}^0u_{ki}^0$ are included in
Eq~(\ref{7}).

In a classic paper, Goddard~\cite{Goddard} started
from Hertz contacts between grains, and considered the
structure of the energy and stress. He concluded that,
if the topology of the grain contacts do not change
with stress, the energy is a homogeneous function of
degree $5/2$ in the strain $u_{ij}$, of the form
$w=\Delta^{2.5}\times g(u_s^2/\Delta^2\!, \,
u^0_{ij}u^0_{jk}u^0_{ki}/\Delta^3)$, where $g$ is an
arbitrary function. As Eq~(7) is clearly a special
case of this general energy, we take this as a
further, microscopically founded support for our
starting point.

There is an instructive analogy between the granular
stress-strain relation, Eq~(\ref{8}), and the van der
Waals equation of state for real gases. The Boyle's law
is stable everywhere while the van der Waals equation has
a non-physical zone, the liquid-gas instability, in which
the compressibility is negative. Similarly, the Hooke's
law is stable everywhere, but the granular stress-strain
relation has a forbidden region, that of yield. Note
\begin{equation}\label{13}
  \left.\partial P/\partial
\Delta\right|_{\sigma _s}\geq 0
\end{equation}
is implied by Eqs~(\ref{10},\ref{11}), see appendix, so
this forbidden region is also characterized by a negative
compressibility. The actual innovation of the van der
Waals theory is the fact that the condition for the onset
of the liquid-gas transition, instead of being an extra
input, is implied by the free energy. Similarly, yield is
now a result of elasticity.

\subsection{Granular Stress-Strain Relation}
The granular stress-strain relation, Eq~(\ref{8}), and
the definitions of Eq~(\ref{3}) imply
\begin{eqnarray}
P &=&\Delta ^{3/2}\left({\cal B}+\textstyle\frac 12{\cal
A}u_s^2/\Delta ^2\right), \label{14} \\ \sigma _s
&=&2{\cal A}\Delta ^{1/2}u_s . \label{15}
\end{eqnarray}
Eliminating $\Delta $, we obtain
\begin{equation}
{\cal B}\sigma _s^4-8{\cal A}^3Pu_s^3\sigma _s+8{\cal
A}^5u_s^6=0. \label{16}
\end{equation}
\begin{figure}[t]
\begin{center}
\includegraphics[scale=0.7]{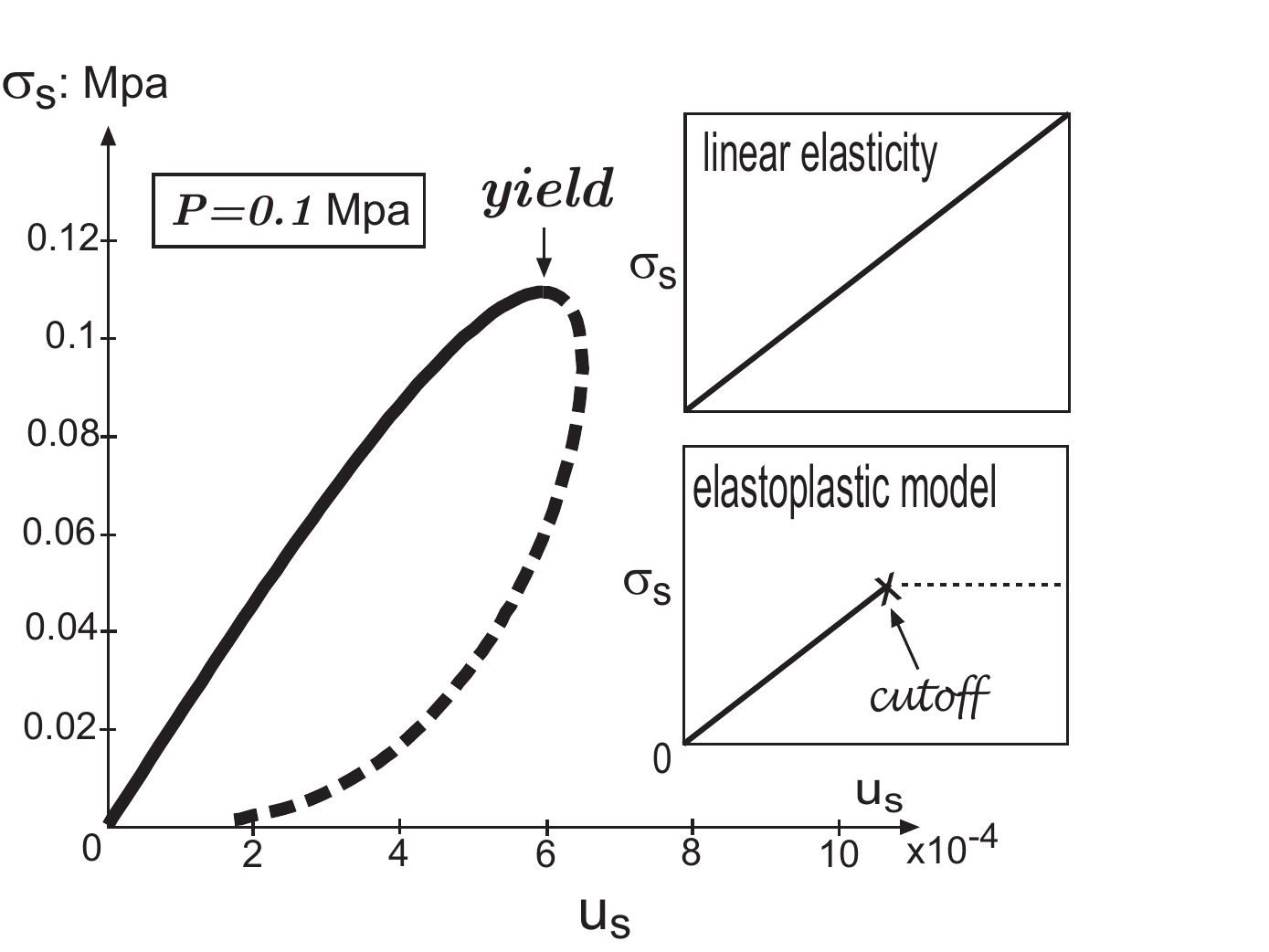}
\end{center}
\caption{Shear stress versus shear strain for given
pressure: for granular elasticity, linear elasticity
(upper insert), and elastoplastic theory (lower insert).}
\label{fig1}
\end{figure}
Fig.~\ref {fig1} plots $\sigma _s$ versus $u_s$ for the
fixed pressure of $P=0.1$ Mpa. Note how remarkably linear
the plot is -- almost until yield, where the curve turns
back abruptly. (Dashed lines are used throughout for
unstable states.) This behavior is approximated by the
elastoplastic model, frequently used in soil mechanics:
Linear elasticity followed by yield and flat plastic
motion, see the lower inserts in Fig.~\ref {fig1}.
Nonlinearity is relevant only when yield is close.

\begin{figure}[tbp]
\begin{center}
\includegraphics[scale=0.7]{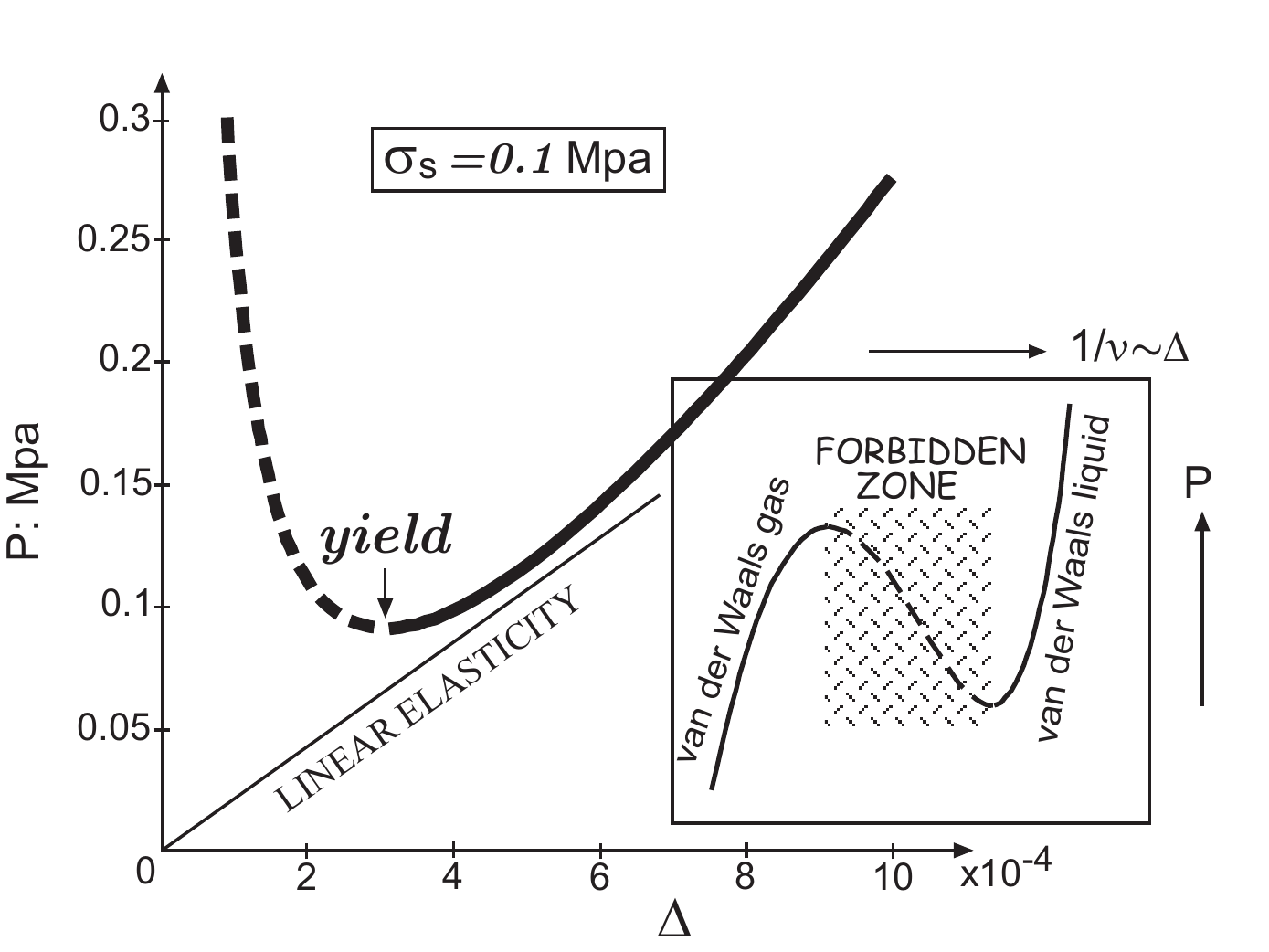}
\end{center}
\caption{Thick line: Pressure versus compression at fixed
shear. Dashed lines represent unstable states. Thin
straight line: The same curve for linear elasticity.
Insert: The analogous instability in the isothermal curve
of the van der Waals equation of state.} \label{fig2}
\end{figure}
If instead $u_s$ is eliminated from
Eqs(\ref{14},\ref{15}), the expression
\begin{equation}
\sigma _s^2+8{\cal AB}\Delta ^3-8{\cal A}P\Delta^{3/2}=0
\label{17}
\end{equation}
allows a plot of pressure $P$ versus compression
$\Delta$, at given $\sigma _s=0.1$ Mpa, see Fig.~\ref
{fig2}. The pressure increases with the compression,
implying a positive compressibility, only in the region
of large $\Delta $. The compressibility is negative where
$\Delta $ is small, and the stability condition,
Eq~(\ref{9}) or (\ref{13}), is violated. The van der
Waals equation of state, $\left( P-a/v^2\right)
(v-b)=RT$,  is quite similar, where $1/v$ corresponds to
$\Delta$, $R$ is the gas constant and $v$ the molar
volume, see eg.~\cite{Callen}. The system can be either
in the dense liquid state or the rarefied gaseous phase,
with the zone in between forbidden, see insert of
Fig.~\ref{fig2}.

\begin{figure}[bp]%
\begin{center}
\includegraphics[scale=0.7]{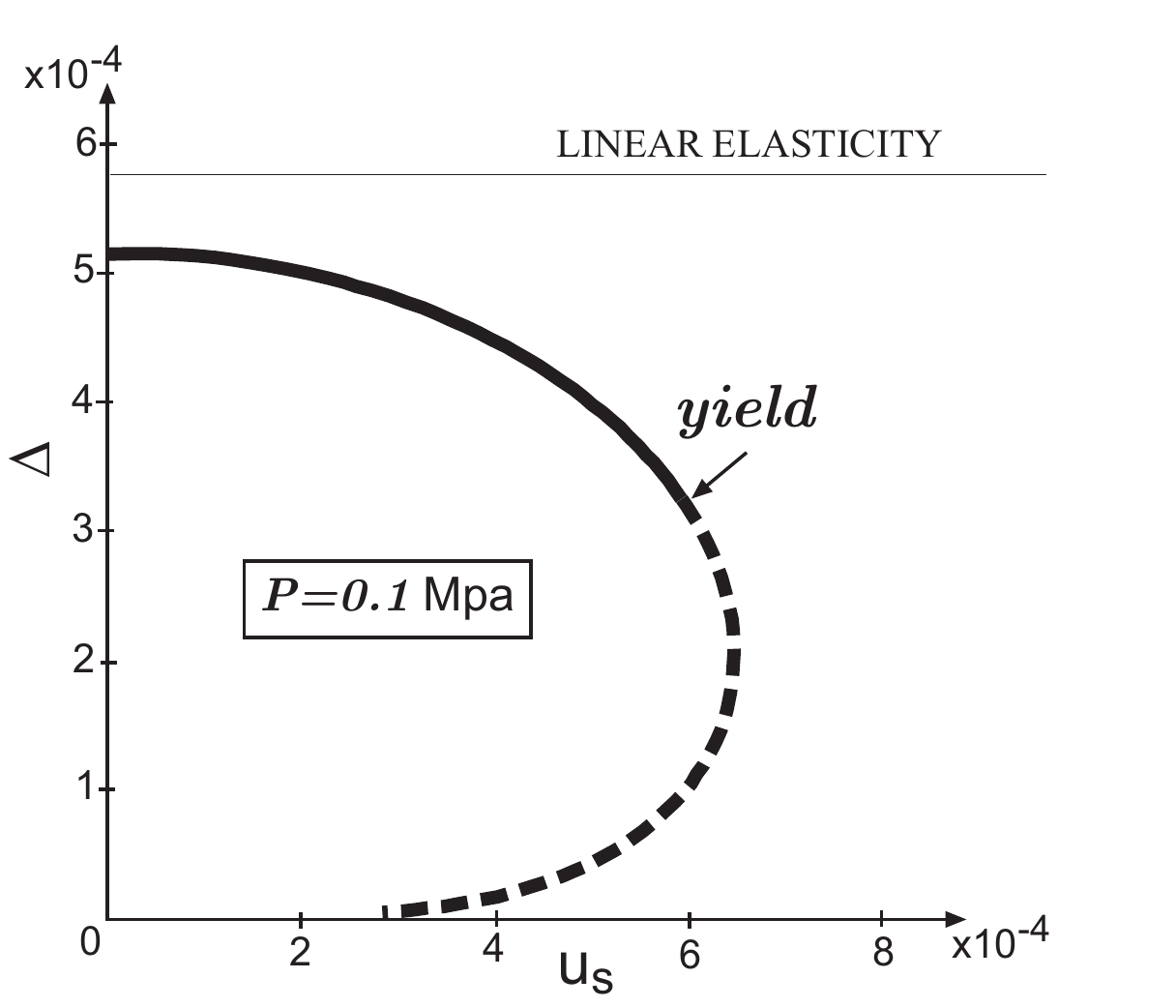}
\end{center}
\caption{Compression $\Delta$ versus shear strain $u_s$,
at fixed pressure. The dashed line is again unstable. In
linear elasticity, the same curve is a horizontal
straight line.} \label{fig3}
\end{figure}
\begin{figure}[t]
\begin{center}
\includegraphics[scale=0.7]{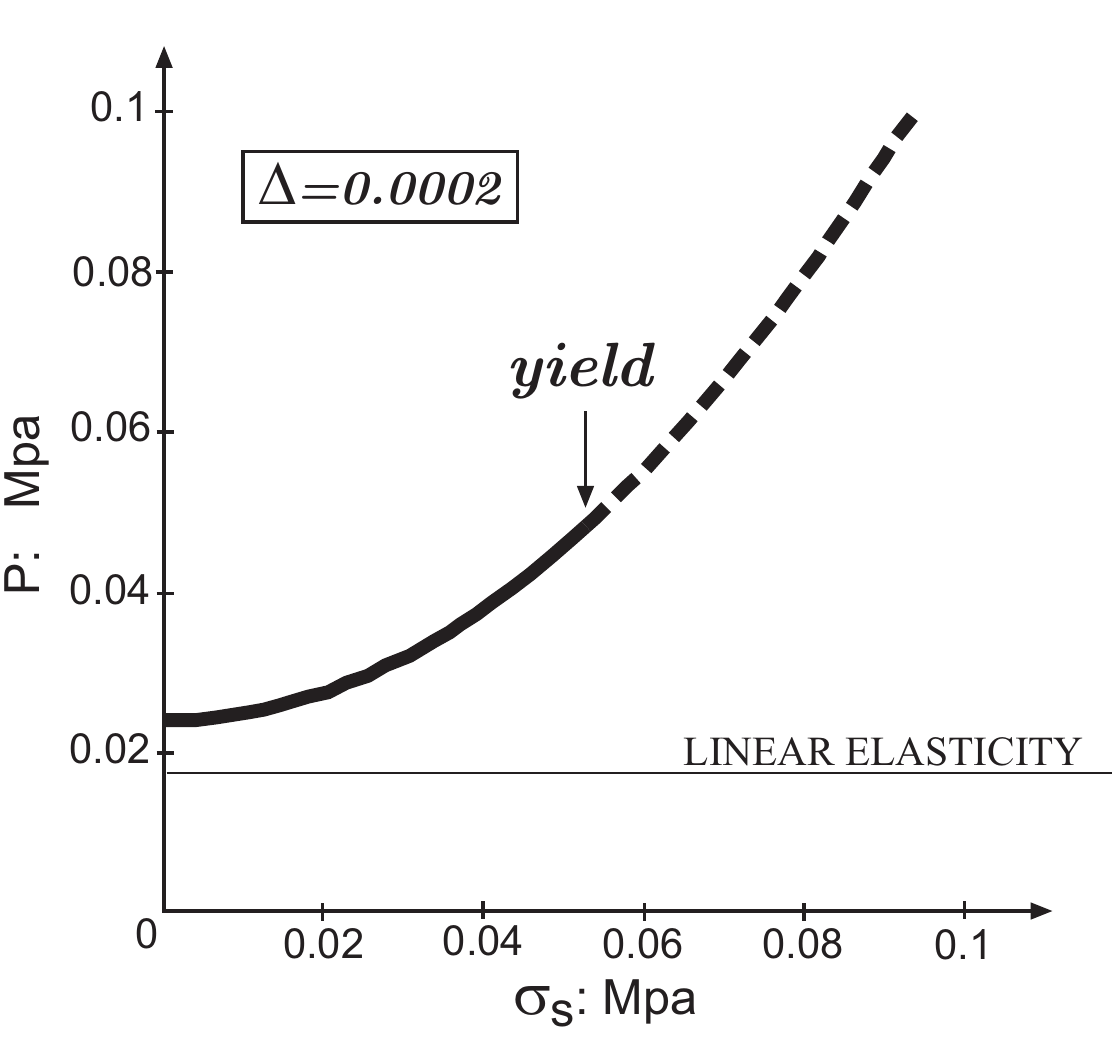}
\end{center}
\caption{Pressure $P$ versus shear stress $\sigma _s$, at
fixed compression. The dashed line is unstable. In linear
elasticity, the same curve is a horizontal straight line.
} \label{fig4}
\end{figure}
Alternatively, we may plot $\Delta$ versus $u_s$ at fixed
$P$, or $P$ versus $\sigma _s$ at fixed $\Delta$, see
Figs.~\ref{fig3} and \ref{fig4}, both showing clear
evidence of ``volume dilatancy," the fact (first noticed
by Reynold) that granular systems expand with shear, or
$\partial\Delta/\partial u_s|_P\not=0$, or $\partial
P/\partial\sigma_s |_\Delta\not=0$. For linear
elasticity, these plots are simply horizontal, and the
derivatives vanish. If the Boussinesq model,
Eq~(\ref{6}), were employed, all four plots would be
indistinguishable from those of linear elasticity. So the
last term of Eq~(\ref{8}) is indeed essential. (Plastic
motion, not considered here, contribute to additional
dilatancy, and may dominate.)

\subsection{Shear-Dependence of the Elastic Moduli}
\begin{figure}[tbp]
\begin{center}
\includegraphics[scale=0.7]{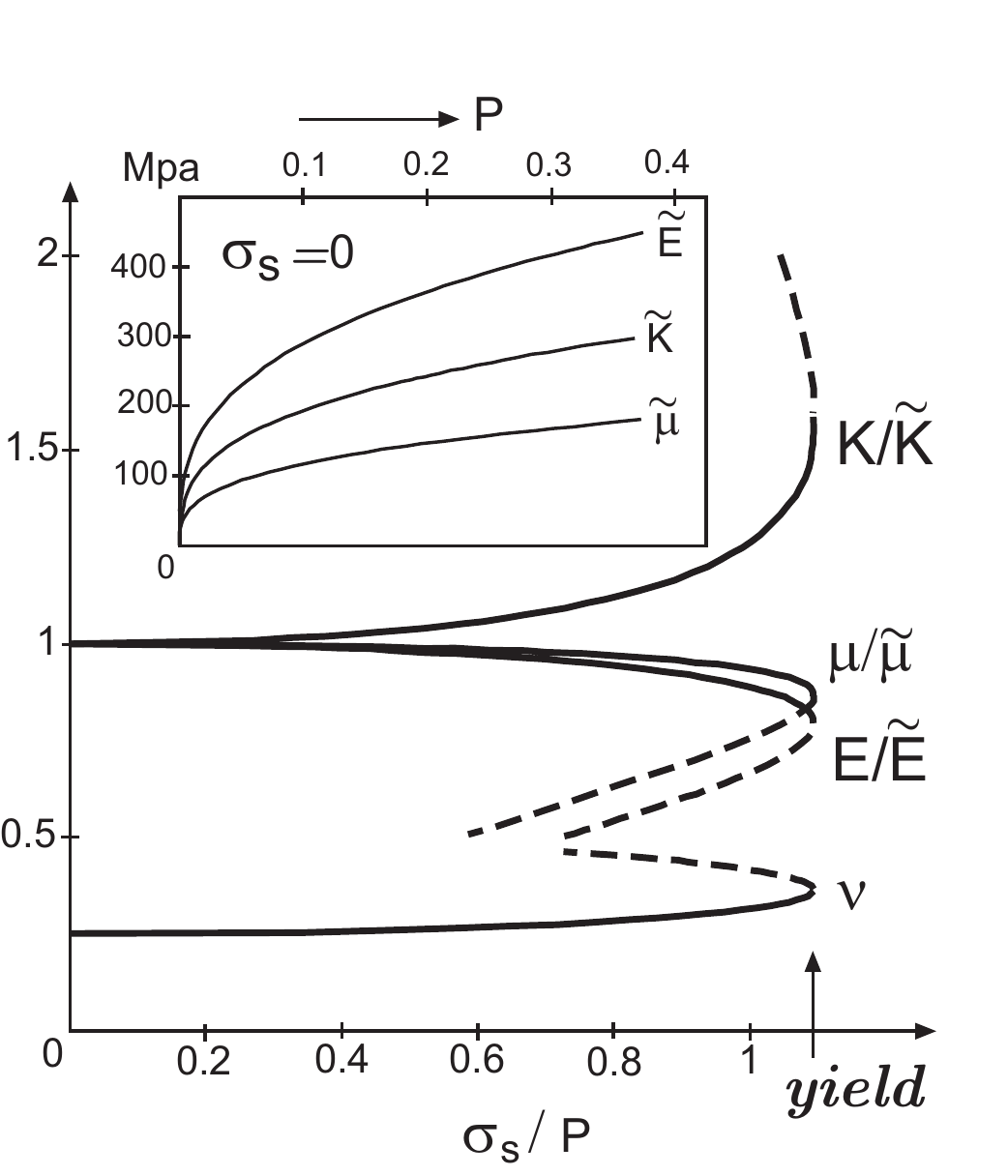}
\end{center}
\caption{Variations of $K,\mu,E,\nu$ with $\sigma_s/P$.
The moduli are rescaled by their values at $\sigma _s=0$,
denoted respectively with a twiddle. Their variation
$\sim P^{1/3}$ is shown in the insert.} \label{fig5}
\end{figure}

The Hooke's law, Eq~(\ref{2}), $\sigma _{ij}=K\Delta
\delta _{ij}-2\mu u_{ij}^0$, may be written as
\begin{equation}
u_{ij}=\frac \nu E\sigma _{nn}\delta _{ij}-\frac{\sigma
_{ij}}{2\mu }, \label{18}
\end{equation}
with the Poisson ratio $\nu $ and the Young modulus $E$
given as
\begin{equation}
E =\frac{9\mu K}{3K+\mu }, \quad \nu =\frac{3K-2\mu
}{6K+2\mu }.  \label{19}
\end{equation}
Requiring the granular stress-strain relation
Eq~(\ref{8}) to assume these familiar forms, either
Eq~(\ref{2}) or (\ref{18}), leads to strain-dependency of
$K,\mu$,
\begin{eqnarray}
K &=&\Delta ^{1/2}\left({\cal B}+\textstyle{\frac 12}
{\cal A}u_s^2/\Delta ^2\right), \label{20}\\ \mu &=&{\cal
A}\Delta ^{1/2},  \label{21}
\end{eqnarray}
and via Eq~(\ref{19}) also of $E,\nu$. As this is an
intuitive way to characterize nonlinear elastic behavior,
we shall consider their shear and pressure dependency
more closely here. Using Eqs~(\ref{14},\ref{15}), we
write these moduli as
\begin{eqnarray}
\mu &=&\widetilde{\mu }\xi ^{1/3}, \ \ \ \ \ \ \ \
K=\widetilde{K}\xi ^{-2/3}, \nonumber \\ E
&=&\widetilde{E}\frac{3{\cal B}+{\cal A}}{3{\cal B}+{\cal
A}\xi }\xi ^{\frac 13},\ \ \nu =\frac{3{\cal B}-2{\cal
A}\xi }{6{\cal B}+2{\cal A}\xi }; \label{22}
\end{eqnarray}
where $\xi$ quantifies shear,
\begin{equation}
\xi =\textstyle\frac 12\left[ 1\pm \sqrt{1-\left( {\cal
B}/2{\cal A}\right) \left( \sigma _s/P\right) ^2}\right],
\qquad\qquad\label{24}
\end{equation}
and $\widetilde{\mu }$, $\widetilde{K}$, $\widetilde{E}$,
$\widetilde{\nu}$ denote the respective value without
shear, at $\xi=1$,
\begin{equation}\widetilde{\mu }={\cal A}\left( \frac P{\cal
B}\right)^{\frac 13},\ \ \widetilde{K} ={\cal B}\left(
\frac P{\cal B}\right) ^{\frac 13},\ \
\widetilde{E}=\frac{9{\cal AB}}{3{\cal B}+{\cal A}}
\left( \frac P{\cal B}\right) ^{\frac 13},  \label{23}
\end{equation}
see Fig.~\ref{fig5}. (The positive sign in Eq~(\ref{24})
is the stable branch, which meets the unstable branch
with the negative sign at yield, where the square root
vanishes.)

As mentioned in the introduction, the
$P^{1/3}$-dependence of the twiddled letters is
well-known. For typical granular behavior, however, the
more relevant dependence is that on shear, which derives
-- same as yield and dilatancy -- from the last term of
Eq~(\ref{8}).

\subsection{The Compliance Tensor}
\subsubsection{Theoretical Expressions} Starting
from Eq~(\ref{8}), the tensor $M_{ijk\ell}$ of
Eq~(\ref{4}) is calculated as
\begin{eqnarray}
M_{ijkl} ={\cal A}\sqrt\Delta\,[({u_s^2}/{4\Delta
^2}+4/3- {3{\cal B}}/{2{\cal A}}) \delta _{ij}\delta
_{kl} \nonumber
\\ -\delta _{ik}\delta _{jl}-\delta _{il}\delta
_{jk}+(u_{ij}\delta _{kl}+\delta _{ij}u_{kl})/\Delta].
\label{25}
\end{eqnarray}
The compliance tensor $\lambda _{ijk\ell}$, defined via
\begin{equation}
\text{d}u_{ij}=\lambda _{ijk\ell}\text{d}\sigma _{k\ell},
\label{26}
\end{equation}
is obtained by inverting $M_{ijk\ell}$,
\begin{eqnarray}\nonumber
\lambda _{ijk\ell} &=&\frac{\left[{\cal A}u_s^2+2 \left(
{\cal A}-{\cal B}\right) \Delta ^2\right] \delta
_{k\ell}\delta _{ij}}{6{\cal A}\Delta ^{1/2} \left( {\cal
A}u_s^2-2{\cal B}\Delta ^2\right) }- \frac{\delta
_{ik}\delta _{j\ell}+\delta _{i\ell}\delta _{jk}}{4{\cal
A}\Delta ^{1/2}} \\ &&+\frac{u_{ij}\Delta \delta
_{k\ell}+u_{k\ell}\Delta \delta _{ij}+u_{ij}u_{k\ell}}{
3\Delta ^{1/2}\left({\cal A}u_s^2-2{\cal B}\Delta
^2\right) },\\  &=&\frac{9{\cal A}^5\sigma _s^2+8\left(
4{\cal A}-9{\cal B}\right) \mu ^6}{54\mu \left( {\cal
A}^5\sigma _s^2-8\mu ^6{\cal B}\right) }\delta
_{k\ell}\delta _{ij}-\frac{\delta _{ik}\delta
_{j\ell}+\delta _{i\ell}\delta _{jk}}{4\mu } \nonumber
\\ &&-\frac{4{\cal A}^3\mu ^3\left( \sigma
_{ij}^0\delta _{k\ell}+\sigma _{k\ell}^0\delta
_{ij}\right) -3{\cal A}^5\sigma _{ij}^0\sigma
_{k\ell}^0}{9\mu \left( {\cal A}^5\sigma _s^2-8\mu
^6{\cal B}\right)}.  \label{27}
\end{eqnarray}
In the first expression $\lambda _{ijk\ell}$ is strain-,
in the second stress-dependent --- where the conversion
is calculated using $\Delta =\mu ^2/{\cal A}^2$,
$u_{ij}^0=-\frac12\sigma _{ij}^0/\mu$, $u_s=\frac12\sigma
_s/\mu$, with $\mu ={\cal A}(\xi P/{\cal B})^{1/3}$, cf.
Eqs~(\ref{22},\ref {23}). The second expression -- a
surprisingly complicated one if the starting expression
for the energy serves as a benchmark -- is what may be
compared to experiments directly.

Before we do this, it is useful to pause and notice that
the last term of both Eq~(\ref{25}) and (\ref{27})
deviates structurally from the isotropic form of
Eq~(\ref{5}). More generally, for an isotropic medium and
in the presence of pure compression ($\sigma^0_{ij}=0,\,
P\not=0$), we may (quite independent of the specific form
of the elastic energy) take $\lambda _{ijk\ell}$ to be
\begin{equation}\label{lambda0}
\lambda^0_{ijk\ell}=\lambda_1\delta_{ij}\delta_{k\ell}
+\lambda_2(\delta_{ik}\delta_{j\ell}
+\delta_{i\ell}\delta_{jk}),
\end{equation}
with $\lambda_1,\lambda_2$ arbitrary scalar functions of
$\Delta,u_s$, and the Lode parameter $L$. This is because
\textbullet~both $\sigma_{ij}$ and $u_{k\ell}$ are
symmetric, hence $\lambda_{ijk\ell}=\lambda
_{jik\ell}=\lambda _{ij\ell k}$; \textbullet~the Maxwell
relation holds, $\partial ^2w/\partial u_{ij}\partial
u_{lk}=\partial ^2w/\partial u_{lk}\partial u_{ij}$,
hence $\lambda _{ijk\ell}=\lambda _{k\ell ij}$. In the
presence of shear, $\sigma^0_{ij}\not=0$,  $\lambda
_{ijk\ell}$ can take on many more terms. To linear order
in $\sigma^0_{ij}$, these are
\[\lambda_3(\sigma^0_{ij}\delta_{k\ell}+\delta_{ij}\sigma^0_{k\ell})+
\lambda_4(\sigma^0_{ik}\delta_{j\ell}+\sigma^0_{i\ell}\delta_{jk}
+\sigma^0_{j\ell}\delta_{ik}+\sigma^0_{jk}\delta_{i\ell}).\]
To second order, we may substitute all above
$\sigma^0_{ij}$ with $\sigma^0_{ik}\sigma^0_{kj}$, and
also add the terms: $\sigma^0_{ij}\sigma^0_{k\ell}$ and
$\sigma^0_{ik}\sigma^0_{j\ell}+
\sigma^0_{jk}\sigma^0_{i\ell}$. We shall refer to
$\lambda^0_{ijk\ell}$ as being isotropic, and the
$\sigma^0_{ij}$-dependent ones as displaying
``shear-induced anisotropy." If the medium were
inherently anisotropic, say because the grains are
pressed into some quasi-periodic array, leading to a
preferred direction $\bf{n}$, the above expression is
more complicated, because $\delta_{ij}$ in
Eq~(\ref{lambda0}) may now be substituted by three
different tensors: $\delta_{ij}-n_in_j$, $n_in_j$, and
$\epsilon_{ijk}n_k$. For triclinic symmetry and without
the Maxwell relation, all 36 elements of
$\lambda_{ijk\ell}$ are independent -- even in the
absence of shear. As mentioned, this ``fabric anisotropy"
is not included in the present consideration, because the
starting expression for the energy, Eq~(\ref{7}), is
isotropic.

\subsubsection{Comparison with Experiments}

Because $\sigma _{ij}$ and $u_{ij}$ are symmetric, each
characterized by six independent components,
Eq~(\ref{26}) may be written as a vector equation,
$\text{d} \vec u=\hat\lambda\text{d}\vec\sigma$, with
$\hat\lambda$ a 6x6 matrix, and $\text{d}u,
\text{d}\sigma$ given as in Eq~(\ref{28}). In the
so-called ``principle system" of coordinates, in which
$\sigma _{ij}$ is diagonal (but not $\delta\sigma
_{ij}$), Kuwano and Jardine take this vector equation to
be given as~\cite{Kuwano-Jardine}
\begin{equation}
\left(
\begin{array}{l}
du_{11} \\ du_{22} \\ du_{33} \\ 2du_{23} \\ 2du_{13}
\\ 2du_{12}
\end{array}
\!\!\right) =\left(
\begin{array}{cccccc}
&  &  & 0 & 0 & 0 \\ & \hat{C} &  & 0 & 0 & 0 \\ & & & 0
& 0 & 0 \\ 0 & 0 & 0 & G_{23}^{-1} & 0 & 0
\\ 0 & 0 & 0 & 0 & G_{13}^{-1} & 0 \\ 0 & 0
& 0 & 0 & 0 & 2G_{12}^{-1}
\end{array}
\!\!\right)\!\! \left(
\begin{array}{l}
d\sigma _{11} \\ d\sigma _{22} \\ d\sigma _{33} \\
-d\sigma _{23} \\ -d\sigma _{13} \\ -d\sigma _{12}
\end{array}
\!\!\right)  \label{28}
\end{equation}
with
\begin{eqnarray}
\hat{C}=\left(
\begin{array}{ccc}
{-1}/{E_1}\, & {\nu _{12}}/{E_2}\, & {\nu _{13}}/{E_3}
\\ {\nu _{21}}/{E_1}\, & {-1}/{E_2}\, & {\nu
_{23}}/{E_3} \\ {\nu _{31}}/{E_1}\, & {\nu _{32}}/{E_2}\,
& {-1}/{E_3}
\end{array}
\right).\label{29}
\end{eqnarray}
$G_{ij}$ is referred to as the shear modulus in the $i-j$
plane, $E_i$ the Young modulus along $i$, and $\nu_{ij}$
the Poisson ratio for ``the effect of the $i$-strain on
$j$-strain." Identifying these moduli with components of
the $\lambda_{ijk\ell}$ tensor,
\begin{eqnarray}\nonumber
 G_{ij}={-1}/{4\lambda _{ijij}},\\\nonumber
E_i={-1}/{\lambda _{iiii}},\\ \nu _{ij}=-{\lambda
_{iijj}}/{\lambda _{jjjj}}\label{30}
\end{eqnarray}
(for $i\neq j$ and without summation over $i$ or $j$), we
may employ Eq~(\ref{27}) to obtain
\begin{eqnarray}
G_{13}=G_{23}=G_{12}=\mu, \qquad\qquad\quad \label{31}
\\ E_i =\frac{27\mu \left( {\cal A}^5\sigma _s^2-8\mu
^6{\cal B}\right)}{9{\cal A}^5\sigma _s^2-72\mu ^6{\cal
B}-{\cal A}s_i^2},\qquad \label{32}
\\ \nu _{ij} =\frac 12\frac{9{\cal A}^5\sigma _s^2-72\mu
^6{\cal B}+2{\cal A}s_is_j}{9{\cal A}^5\sigma _s^2-72\mu
^6{\cal B}-{\cal A}s_j^2}, \label{33}
\end{eqnarray}
with $\mu ={\cal A}(\xi P/{\cal B})^{1/3}$, $s_i\equiv
3{\cal A}^2\sigma _i^0-4\mu ^3$,  $\sigma
_i^0\equiv\sigma _i-P$, and $\sigma_i$ denoting the three
diagonal components of $\sigma_{ij}$ in the principle
system. Before embarking on a comparison, we shall first
establish a few qualitative features from theory:
\textbullet~Without shear, $\sigma _i^0\rightarrow 0$,
all $E_i$ are equal,
\begin{equation}
E_i\rightarrow E_{\sec }= \frac{27{\cal A}{\cal
B}}{2{\cal A}+9{\cal B}}\left( \frac P{\cal B}\right)
^{\frac 13}, \label{34}
\end{equation}
where $E_{\sec }$ is called the secant Young modulus.
Same holds for the Poisson ratios,
\begin{equation}
\nu _{ij}\rightarrow \widetilde{\nu }^{*}= \frac
12\frac{9{\cal B}-4{\cal A}}{9{\cal B}+2{\cal A}}.
\label{35}
\end{equation}
(Note $\widetilde{ \nu }^{*}$ differs from
$\widetilde{\nu }$, and $E_{\sec }$ from $\widetilde{E}$,
by a constant factor.) \textbullet~ Because of
Eq~(\ref{lambda0}) and irrespective of the energy
specified, we have $E_1=E_2=E_3$,
$\nu_{12}=\nu_{13}=\nu_{23}$, and $G_{12}=G_{13}=G_{23}$
in the absence of shear, $\sigma^0_{ij}=0$. Any
discrepancy with experiment therefore implies fabric
anisotropy. \textbullet~Finite shear will split $E_i$ and
$\nu _{ij}$, but not $G_{ij}$, cf. Eq~(\ref{31}) ---
though this is an energy-related feature.
\textbullet~Because of the Maxwell relation, the matrix
$\hat\lambda$ of Eq~(\ref {28}) is symmetric, implying
especially (no summation)
\begin{equation}
\nu _{ij}E_i=\nu _{ji}E_j.  \label{36}
\end{equation}
This symmetry was noted by Love (1927) and adopted by
Kuwano and Jardine in interpreting their
data~\cite{Kuwano-Jardine}. \textbullet~The moduli $E,
\mu, \nu $ are related as $E=2\mu \left( \nu +1\right) $,
see Eqs.(\ref{19}). A similar relation holds for $\mu$,
$E_i$, $\nu _{ik}$ [no summation, see
Eqs~(\ref{32},\ref{33})],
\begin{equation}
E_i\left( 6\mu \nu _{ij}-E_j\right) ^2=4E_j\left( 3\mu
-E_i\right)
 \left(3\mu -E_j\right)  \label{37}
\end{equation}

\begin{figure}[tbp]
\begin{center}
\includegraphics[scale=0.35]{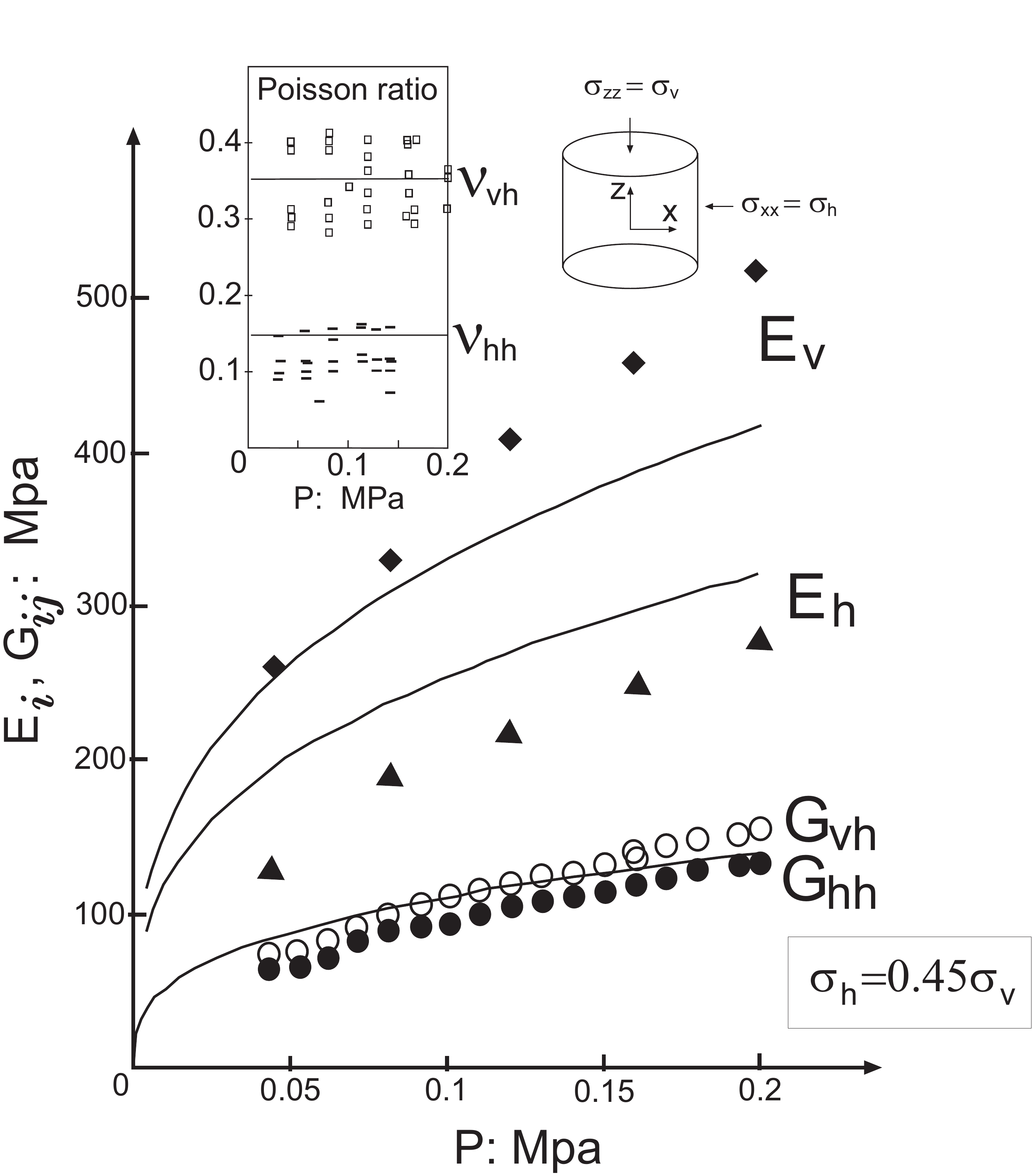}
\end{center}
\caption{Variation with pressure $P$ of the shear moduli
$G_{vh}, G_{hh}$, Young moduli $E_{v}, E_{h}$ and Poisson
ratios $\nu_{vh}$, $\nu_{hh}$ (insert), at $\sigma
_h/\sigma _v=0.45$. Symbols are the same data on Ham
River sand, at a void ratio of 0.66 by Kuwano \& Jardine,
(2002).} \label{fig6}
\end{figure}
It is important to realize that all formulas of this
section hold not only for Cartesian coordinates, $i\to
x,y,z $, but also for cylindrical ones, $i\to
z,\rho,\varphi$. Taking
$\Delta=u_{\varphi\varphi}+u_{\rho\rho}+u_{zz}$, and
similarly for $u_s$, we may again start from the same
energy, Eq~(\ref{7}), and derive all the results here.
[Spatial differentiation is what mars the similarity.
Yet once the strain components $u_{\rho\rho},
u_{\rho\varphi}\dots$ are given, no spatial
differentiation is needed.] The one difference is, for
any constant $\sigma_{ij}$ in Cartesian coordinates,
there is always a principle system. In cylindrical
coordinates, this holds only if the stress is also
cylindrically symmetric. In other words, only if the
stress is uniaxially diagonal,
$\sigma_{ij}=\text{diag} (\sigma_1, \sigma_2,
\sigma_3)$ with $\sigma_2=\sigma_1$ in Cartesian
coordinates, will it be diagonal cylindrically.

Because  Kuwano and Jardine~\cite{Kuwano-Jardine} used an
axialsymmetric device for their measurements, the stress
they apply is indeed cylindrically symmetric, with:
$G_{\rho z}=G_{\varphi z}$, $E_\rho=E_\varphi$, $\nu
_{\rho z}=\nu _{\varphi z}$, $\nu _{z\rho}=\nu
_{z\varphi}$, cf. Eqs.(\ref{31}-\ref{33}) noting
$s_\rho=s_\varphi$. In addition, Eq~(\ref{36}) leads to
$\nu _{\rho\varphi}=\nu _{\varphi\rho}$. Following them,
we refer to the response coefficients being measured as:
$G_{hh}\equiv G_{\rho\varphi}$, $G_{vh}\equiv G_{\rho
z}=G_{\varphi z}$, $E_h\equiv E_\rho=E_\varphi$,
$E_v\equiv E_z$, $\nu_{hh}\equiv \nu _{\rho\varphi}=\nu
_{\varphi\rho}$, $\nu _{hv}\equiv\nu _{\rho z}=\nu
_{\varphi z}$, $\nu _{vh}\equiv \nu _{z\rho}=\nu
_{z\varphi}$, where $h$ is the horizontal directions,
either $\rho$ or $\varphi$, and $v$ the vertical
direction $z$, see the cylinder of Fig.~\ref{fig6}. The
main plots of Fig.~\ref{fig6} compare the theoretical
curve [calculated by taking $\sigma _\rho=\sigma
_\varphi=\sigma _h$ and $\sigma _z=\sigma _v$ in
Eqs.(\ref{31}-\ref{33})] and the experimental data
[measured with Ham River sand] of $E_h$, $E_v$, $G_{vh}$,
$G_{hh}$, as functions of $P$, for $\sigma _h=0.45\,
\sigma _v$. The insert shows the same comparison for $\nu
_{vh},\nu _{hh}$. We especially note that theory and
experiment agree on the ordering of the induced
anisotropy, ie $\nu _{vh}>\nu _{hh}$, $E_v>E_h$ and
$G_{hh}\approx G_{vh}$, which are pairwise equal in
linear elasticity and the Boussinesq model. (The slight
difference between $G_{hh}$, $G_{vh}$ is, as mentioned,
the result of fabric anisotropy present in the sample.)
For a theory without any useful fit parameter, the
agreement must be considered a convincing verification of
the elastic approach which, instead of postulating the
stress-dependence of 21 (or even 36) independent
components of $\lambda_{ijkl}$ directly, looks for one
appropriate scalar expression for the energy $w$. Even if
it is heavy-handedly simplified, a large number of
geometric correlation is preserved by the mere fact that
$\lambda_{ijkl}$ is obtained via a double
differentiation. This must be the main reason why the
calculated $\lambda_{ijkl}$ stands up so surprisingly
well when compared to the extensive data
of~\cite{Kuwano-Jardine}.

\begin{figure}[tbp]
\begin{center}
\includegraphics[scale=0.7]{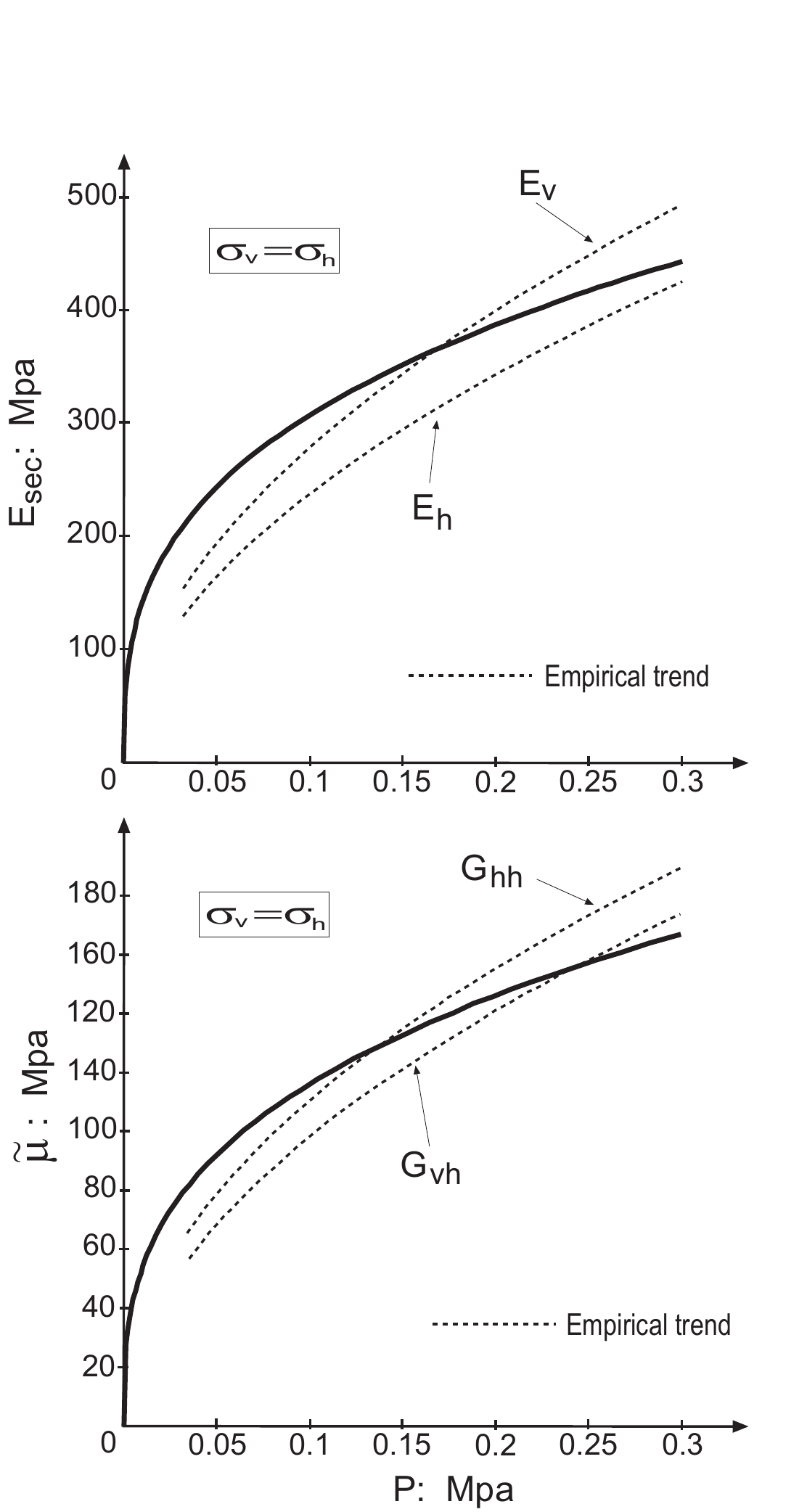}
\end{center}
\caption{Variation of Young and shear moduli, $E_{sec}$
and $\mu$, with pressure $P$, for the case of vanishing
shear, $\sigma _v=\sigma _h$. The dotted lines are the
empirical formula of Kuwano \& Jardine (2002), for the
Ham River sand at the void ratio $e=0.66$. The split is
proof of fabric anisotropy.} \label{fig7}
\end{figure}

Kuwano and Jardine~\cite{Kuwano-Jardine} employ the
following empirical formulas (in Mpa) for the Ham River
sand,
\begin{eqnarray}
E_v &=&204f\left( \sigma _v/P_a\right) ^{0.52}
\label{Ev-exp} \\ E_h &=&174f\left( \sigma _h/P_a\right)
^{0.53}  \label{Eh-exp} \\ G_{vh} &=&72f\left( \sigma
_v/P_a\right) ^{0.32}\left( \sigma _h/P_a\right) ^{0.2}
\label{Gvh-exp} \\ G_{hh} &=&81f\left( \sigma
_v/P_a\right) ^{-0.04}\left( \sigma _h/P_a\right) ^{0.53}
\label{Ghh-exp}
\end{eqnarray}
where $P_a=0.1013$ Mpa is the atmospheric pressure and
$f=(2.17-e)^2/(1+e)$. ($f=1.3736$ for the void ratio
$e=0.66$.) Fig.~\ref{fig7} shows the theoretical and
experimental values for $E_h$, $E_v$, $G_{vh}$ and
$G_{hh}$, as functions of $P$ for the isotropic case
$\sigma _h=\sigma _v$. The fact that $E_h$, $E_v$ and
$G_{vh}$, $G_{hh}$ are pairwise different, indicates (as
discussed above) fabric anisotropy. Moreover, the
theoretical curves are $\sim P^{1/3}$, yet experimental
ones seem to back a larger power: $\sim P^{1/2}$. As
discussed, \textbullet~this is a known contradiction
between Hertz contact and sound data, with possible
explanations provided by Goddard\cite{Goddard} and de
Gennes~\cite{deGennes96}, \textbullet~and a question of
simplicity versus accuracy in the present approach.

\begin{figure}[t]
\begin{center}
\includegraphics[scale=0.53]{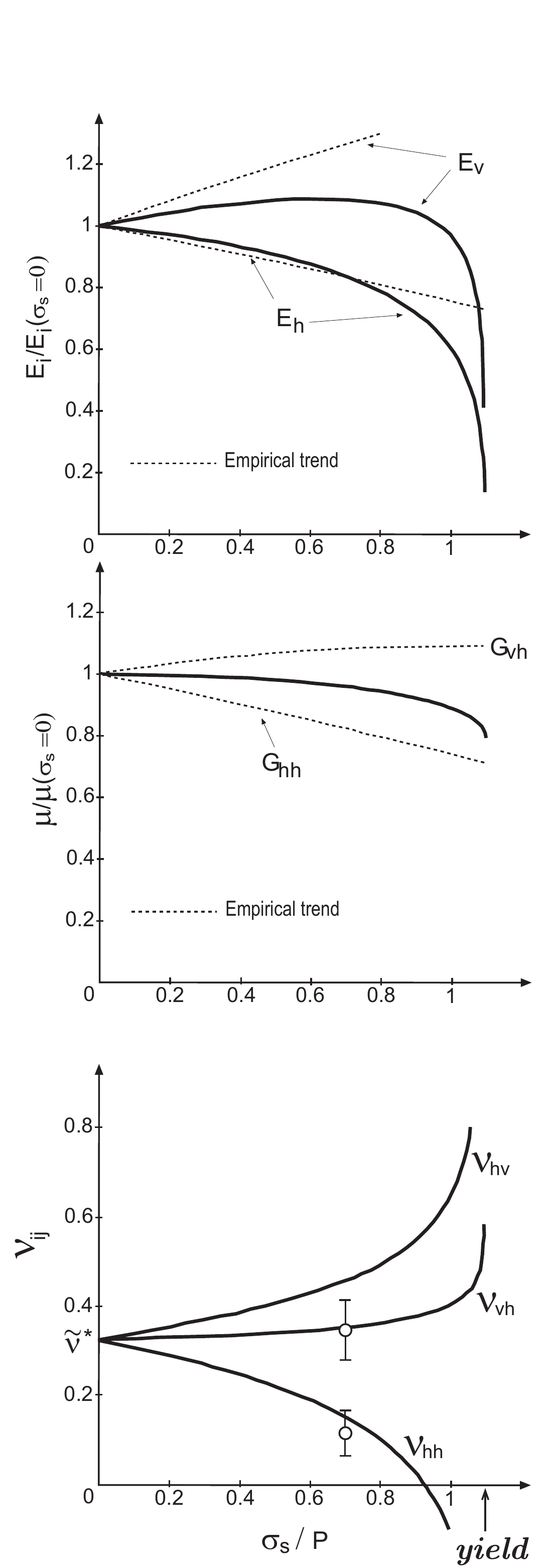}
\end{center}
\caption{Upper, middle, and lower figures show the Young
moduli, shear moduli and Poisson ratios as functions of
$\sigma_s/P$. The dotted lines present the empirical
formulas of Kuwano \& Jardine (2002) for the Ram River
sand , at the void ratio $e=0.66$.} \label{fig8}
\end{figure}

Fig.\ref{fig8} displays the effect of shear on different
moduli, with $\sigma _h\not=\sigma _v$. The upper, middle
and lower figures respectively plot the Young moduli
$E_i$, the shear modulus $\mu $ (both scaled by their
isotropic values, $E_{\rm sec}, \tilde\mu$), and the
Poisson ratios $\nu_{ij}$. In agreement with the
empirical formulas Eqs~(\ref{Ev-exp}-\ref{Ghh-exp}),
$E_v$ increases with $\sigma _s/P$, while $E_h$
decreases, in the region away from yield. As yield is
approached, both drop quickly to zero. This critical,
pre-yield behavior is clearly absent for the empirical
formulas and is of interests for future experiments. In
theory, $G_{vh},G_{hh}$ are equal, decreasing with
$\sigma _s/P$ moderately, by less than 20\%. In
experiments, the shear moduli are split, with one
increasing, the other decreasing. The discrepancy between
the theory and experiment is in the range from $\sigma
_s/P=0$ to $0.6$ within 20\%. This need not be a result
from fabric anisotropy, as a more complicated energy
expression will also do.

Variation of the Poisson ratios $\nu _{vh},\nu _{hv},\nu
_{hh}$ is given by Eq~(\ref{33}). As depicted, $\nu
_{vh}$ and $\nu _{hv}$ increase, while $\nu _{hh}$
decreases, with $\sigma _s/P$, all being divergent at
yield. No empirical formulae for the ratios are given
in~\cite{Kuwano-Jardine}, and the two circles in the plot
simply depict the values from the insert of
Fig~\ref{fig1}. However, $\nu _{hh}=E_h/(2G_{hh})-1$ was
assumed to hold by the authors, and interestingly, it may
be derived by taking $i=h$, $j=h$ in Eq~(\ref{37}),
yielding $\nu _{hh}=E_h/(2\mu )-1$.

Assuming that both coefficients $\cal A,B$ of
Eq~(\ref{7}) are proportional to $f$ of
Eqs~(\ref{Ev-exp}-\ref{Ghh-exp}), agreement between
experiment and theory is extended to all values of the
void ratio. Comparison was also made to Kuwano and
Jardine's data gained using glass
ballotini~\cite{Kuwano-Jardine}. Taking ${\cal A}=4200$,
${\cal B}=\frac53{\cal A}=7000$, we find similar
agreement.

\subsection{The Elastic Part of Flow Rules}

The increment relation, Eq~(\ref{4}), may also be
written in the matrix form $\text{d}\vec{\sigma
}=\hat{M}\text{d}\vec{u}$, with $\hat{M}$ a symmetric
$6\times 6$ matrix, and $\text{d}\vec{\sigma },
\text{d}\vec{u}$ still given as in Eq~(\ref{28}). The
determinant, $\det \hat{M}=9{\cal A}^5\left( 2{\cal
B}\Delta ^2-{\cal A}u_s^2\right) \Delta$, calculated
from Eq~(\ref{25}), vanishes at the yield surface,
${\cal A}u_s^2=2{\cal B}\Delta ^2$, because an
Eigenvalue, call it $m_1$, also does. (This is not a
coincidence as $\hat{M}$ is the Jacobian matrix of the
energy function, which is positive only in the stable
region. It may be of interest to note that the
determinant of the Bousinesq model,
$\det\hat{M}=9{\cal A}^5 \left(3{\cal B}+4{\cal
A}\right) \Delta^3$, never vanishes.) The associated
Eigenvector $\vec{m}_1$ points along the direction at
which a finite deformation $\text{d}\vec{u}\neq 0$ may
take place under constant stress $\text{d}\sigma
_{ij}=0$. We refer to $\vec{m}_1$ as the elastic flow
direction, since $\vec{m}_1\|\text{d}\vec{u}$ is only
the elastic contribution of the strain. Setting
$\text{d}\sigma_{ij}=0$ in Eq~(\ref{4}) and using
${\cal A}u_s^2=2{\cal B}\Delta ^2$, we obtain
\[\text{d}u_{ij}=-\frac 12\left( \delta _{ij}+
\frac{u_{ij}}\Delta \right)\text{d}\Delta =\left(
\sqrt{\frac {\cal B}{2{\cal A}}}\frac{\sigma
_{ij}^0}{\sigma _s}-\frac{\delta _{ij}
}3\right)\text{d}\Delta.\]
The calculated $\text{d}u_{ij}\to \text{d}\vec u$ is
the Eigenvector $\vec m_1$. Remarkably, one can
rewrite this equation as
$\text{d}u_{ij}/\text{d}\Delta =\partial g/\partial
\sigma _{ij}$, or
\begin{equation}
\vec{m}_1\parallel \partial g/\partial \vec{\sigma},
\quad \text{with}\quad g=\sqrt{{\cal B}/2{\cal A}}\sigma
_s-P, \label{flow-potential}
\end{equation}
implying that the elastic flow direction is
perpendicular to the yield surface, as defined by the
equation $g=0$. If the plastic contribution to the
strain field may be neglected for some reasons, this
property is referred to as the \textit{associated flow
rule} see~\cite{wroth,Huang}.

\appendix
\section{energetic stability}

In the main text, we considered the convexity of the
energy with respect to the variables $u_s$ and $\Delta$.
Relevant is the convexity with respect to $u_{ij}$. As
the transformation between these two sets of variables is
nonlinear, we bear the burden of proof that both are
equivalent.

Thermodynamic stability requires the elastic energy to be
a convex function of its six strain variables, or linear
combinations of them. This means that all eigenvalues of
the Jacobian matrix $\partial ^2w/\partial X_\alpha
\partial X_\beta $ are positive. We take:
$X_1=u_{xy},X_2=u_{xz},X_3=u_{yz}$, $X_4=\left(
u_{xx}-u_{zz}\right) /2$, $X_5=\left(
u_{xx}-2u_{yy}+u_{zz}\right) /(2\sqrt{3})$,
$X_6=-u_{xx}-u_{yy}-u_{zz}$, with $Q=u_s^2=2\sum_{\alpha
=1}^5X_\alpha ^2$. For an energy of the form $w=w(\Delta
,Q)=w(X_6,Q)$ and denoting $f\equiv 4\partial w/\partial
Q$, $a\equiv
\partial ^2w/\partial \Delta ^2$,
$b\equiv4\partial ^2w/\partial Q\partial \Delta $,
$c=16\partial ^2w/\partial Q^2$, the Jacobian matrix is
\[
\left(
\begin{array}{llllll}
f+cX_1^2 & cX_1X_2 & cX_1X_3 & cX_1X_4 & cX_1X_5 & bX_1
\\ cX_1X_2 & f+cX_2^2 & cX_2X_3 & cX_2X_4 & cX_2X_5 &
bX_2 \\ cX_1X_3 & cX_2X_3 & f+cX_3^2 & cX_3X_4 & cX_3X_5
& bX_3 \\ cX_1X_4 & cX_2X_4 & cX_3X_4 & f+cX_4^2 &
cX_4X_5 & bX_4 \\ cX_1X_5 & cX_2X_5 & cX_3X_5 & cX_4X_5 &
f+cX_5^2 & bX_5 \\ bX_1 & bX_2 & bX_3 & bX_4 & bX_5 & a
\end{array}
\!\!\right)
\]
with its six eigenvalues given as $f_{1-4}=f$ and
\begin{equation}
f_{\pm }=\frac{f+a}2+\frac{cQ}4\pm \frac 12\sqrt{\left(
f-a+\frac{cQ} 2\right) ^2+2b^2Q}. \label{a1}
\end{equation}
They are all positive if, and only if, $f>0$,
$2af+acQ-b^2Q>0$, $f+a+cQ/2>0$, or equivalently,
\begin{eqnarray}
\frac{\partial w}{\partial Q} >0,\qquad 4\frac{\partial
w}{\partial Q}+\frac{\partial ^2w}{\partial \Delta ^2}+8Q
\frac{\partial ^2w}{\partial Q^2} >0.   \label{a2}
\\ \frac{\partial ^2w}{\partial \Delta
^2}\frac{\partial w}{\partial Q}+2Q\frac{\partial
^2w}{\partial Q^2}\frac{\partial ^2w}{\partial \Delta
^2}-2Q\left( \frac{\partial ^2w}{\partial Q\partial
\Delta }\right) ^2 >0,  \label{a3}
\end{eqnarray}
Because $u_s^2=Q$, or $2u_s\left( \partial w/\partial
Q\right) =\partial w/\partial u_s$, $2u_s\left(
\partial ^2w/\partial \Delta
\partial Q\right) =\partial ^2w/\partial
\Delta \partial u_s$, $4u_sQ\times \left(
\partial ^2w/\partial Q^2\right) =u_s\left( \partial
^2w/\partial u_s^2\right) -\partial w/
\partial u_s$, these
conditions are equivalent to Eqs~(\ref{10},\ref{11}), or
\begin{eqnarray}
&&\frac{\partial w}{\partial \Delta }>0,\ \frac{\partial
^2w}{\partial \Delta ^2}>0,\ \frac{\partial ^2w}{\partial
u_s^2}>0,  \label{a5} \\ &&\frac{\partial ^2w}{\partial
\Delta ^2}\frac{\partial ^2w}{\partial u_s^2}
>\left( \frac{\partial ^2w}{\partial u_s\partial \Delta }\right) ^2.
\label{a6}
\end{eqnarray}
For the energy of Eq~(\ref{7}), the inequalities
(\ref{a5}) imply ${\cal A}>0, {\cal B}>0$, while
Eq~(\ref{a6}) gives the yield condition (\ref{12}). Using
$P=\partial w/\partial \Delta $ and $\sigma _s=\partial
w/\partial u_s$ we can also write Eqs~(\ref{a5},\ref{a6})
as
\begin{eqnarray}
\left( \frac{\partial P}{\partial \Delta }\right)
_{u_s}>0,\ \ \left( \frac{\partial \sigma _s}{\partial
u_s}\right) _\Delta >0,  \label{a7}\\
%\end{eqnarray}
%\begin{eqnarray}
\left( \frac{\partial P}{\partial \Delta }\right)
_{u_s}\left( \frac{
\partial \sigma _s}{\partial u_s}\right) _\Delta >
\left( \frac{\partial P}{
\partial u_s}\right) _\Delta ^2  \label{a8}
\end{eqnarray}
The Maxwell relation $\left.\partial P/\partial
u_s\right|_\Delta =\left.\partial \sigma _s/\partial
\Delta \right| _{u_s}$ and the thermodynamic
identities $\left.\partial P/\partial \Delta
\right|_{u_s}=\left.
\partial P/\partial \Delta \right|_{\sigma _s}+\left.
\partial P/\partial \sigma _s\right| _\Delta\cdot \left.
\partial \sigma _s/\partial \Delta \right|_{u_s}$,
$\left.\partial P/\partial u_s\right|_\Delta = \left.
\partial P/\partial \sigma _s\right|_\Delta\cdot \left.
\partial \sigma _s/\partial u_s\right|_\Delta$ , imply
an alternative stability condition,
\begin{equation}
\left( \partial P/\partial \Delta \right) _{\sigma _s}>0.
\label{a9}
\end{equation}

\end{document}